\documentclass[prc,preprint,nofootinbib,preprintnumbers,longbibliography,
superscriptaddress,tightenlines
]{revtex4-1}

\usepackage{amsmath,mathrsfs}    
\usepackage{graphicx}   
\usepackage{subfigure}
\usepackage{color}
\usepackage[colorlinks=true,linkcolor=blue,citecolor=blue, urlcolor=blue]{hyperref}   
\usepackage{bm} 
\usepackage{ulem}
\usepackage[inline]{enumitem}
\usepackage{soul}
\usepackage{graphicx}
\usepackage{amsmath,amssymb}
\usepackage{soul}
\usepackage{textcomp}
\usepackage{hyperref}
\usepackage{subfigure}
\usepackage[toc,page]{appendix}
\usepackage{array}   
\newcolumntype{C}{>{$}c<{$}} 

\usepackage{tikz}

\usepackage{array}   
\newcolumntype{C}{>{$}c<{$}} 

\definecolor{MyDarkBlue}{RGB}{22, 79, 134}
\definecolor{MyLightRed}{RGB}{200, 37, 6}

\def\F{{\mathcal H}}

\def\st{\begin{equation}}
\def\stp{\end{equation}}
\def\bg{\begin{eqnarray}}
\def\nd{\end{eqnarray}}
\def\Eq#1{eq.~(\ref{#1})}

\def\Fig#1{Fig.~\ref{#1}}

\def\Sect#1{Sect.~\ref{#1}}

\def\llangle{\left\langle}
\def\rrangle{\right\rangle}
\def \bes {\begin{subequations}}
\def \ees {\end{subequations}}

\def \F
\def\chemconst{{\chi_0}}

\newcommand{\App}[1]{{App.~\ref{#1}}}

\newcommand{\QCD}{{\scriptscriptstyle \mathrm{\tiny QCD}}}
\newcommand{\hatchi}{\hat{\chi}}
\newcommand{\mm}{{\mathfrak m}}

\begin{document}

\title{Dynamics of the $O(4)$ critical point in QCD}

\author{Adrien Florio}
\email[]{adrien.florio@stonybrook.edu}
\affiliation{Center for Nuclear Theory, Department of Physics and Astronomy, Stony Brook University, New York 11794-3800, USA}
\author{Eduardo Grossi}
\email[]{eduardo.grossi@ipht.fr}
\affiliation{Center for Nuclear Theory, Department of Physics and Astronomy, Stony Brook University, New York 11794-3800, USA}
\affiliation{
Institut de physique th\'{e}orique, Universit\'{e} Paris Saclay, CNRS, CEA, F-91191 Gif-sur-Yvette, France}
\affiliation{CPHT, CNRS, Ecole Polytechnique, IP Paris, F-91128 Palaiseau, France}
\author{Alexander Soloviev}
\email[]{alexander.soloviev@stonybrook.edu}
\affiliation{Center for Nuclear Theory, Department of Physics and Astronomy, Stony Brook University, New York 11794-3800, USA}
\author{Derek Teaney}
\email[]{derek.teaney@stonybrook.edu}
\affiliation{Center for Nuclear Theory, Department of Physics and Astronomy, Stony Brook University, New York 11794-3800, USA}

\date{\today}
   \begin{abstract}
      We perform a real-time simulation of the $O(4)$ critical point of QCD, which lies in the dynamic universality class of ``Model G". The axial charge and the order parameter  $\phi_a =(\sigma, \vec{\pi})$ exhibit a rich dynamical interplay, which reflects the qualitative differences in the hydrodynamic effective theories  above and below $T_c$.  From the axial charge correlators on the critical line we extract a dynamical critical exponent of $\zeta=1.47 \pm 0.01 ({\rm stat})$, which is compatible with  the theoretical expectation of $\zeta = d/2$ (with $d=3$) when systematic errors are taken into account. At low temperatures, we quantitatively match the $O(4)$ simulations to the superfluid effective theory of soft pions.
   \end{abstract}


\maketitle

\clearpage

\section{Introduction}
\label{sec:intro}

Chiral symmetry breaking and the chiral phase transition play a prominent role
in QCD at finite temperature.
In the limit of two massless flavors, the transition from a chirally
restored phase $T>T_c$,  to a chirally broken phase $T<T_c$, is second order and is in the $O(4)$ universality class~\cite{Pisarski:1983ms,Rajagopal:1992qz}.
Although the static properties of the $O(4)$
critical point have been studied in detail both numerically and theoretically~\cite{Toussaint_1997,Kanaya:1994qe,Engels:2011km,Engels:2014bra,ParisenToldin:2003hq,Butti:2003nu,Braun:2009ruy,Braun:2010vd,Braun:2011iz,Braun:2020ada}, the dynamic scaling properties of the  critical region demand additional study, which is the goal of this work.

This study may appear academic: in the real world the finite quark mass explicitly breaks chiral symmetry, reducing the influence of the $O(4)$ critical point
on  the static and dynamic correlators  of QCD at finite temperature.
 However, lattice QCD
computations of the chiral condensate as a function of quark mass show that the
qualitative and, to some extent, even quantitative properties of the chiral
crossover can be understood using static $O(4)$ scaling functions~\cite{Ding:2019prx,Kaczmarek:2020sif}.
These scaling
functions predict the singular behavior of the chiral condensate near the
pseudocritical temperature $T_{\rm pc}$ and other static observables.
Motivated by the lattice effort, we will perform the  real-time simulations of the $O(4)$ critical region, which we hope can provide an analogous understanding of the scaling of dynamic correlators in QCD in the crossover region.

The current study is also motivated by experimental results on the momentum spectra of particles produced during the collisions of heavy ions at the Relativistic Heavy Ion Collider (RHIC) and the Large Hadron Collider (LHC). In much of the accessible momentum range these spectra  are remarkably well described by ordinary viscous hydrodynamics~\cite{Heinz:2013th}.  From a theoretical perspective, the relevant symmetry group of QCD close to the chiral limit is approximately $SU_L(2) \times SU_R(2)$,  leading to the conservation of  iso-vector  charge, and the approximate conservation of the iso-axial vector charge. The corresponding densities should be included as
additional fields in the hydrodynamic description. When chiral symmetry is
spontaneously broken, the pions $\vec{\pi}$ (which are the associated
Goldstone bosons)  should also be added to the hydrodynamic fields, and
the appropriate hydrodynamics resembles a non-abelian
superfluid~\cite{Son:1999pa,Grossi:2020ezz}. Finally, close to the $O(4)$ critical point
where the $\sigma$ meson is also light,  the  $O(4)$ order parameter
$\phi_a \sim (\sigma,\vec{\pi})$  also should be included as an
additional hydrodynamic degree of freedom~\cite{Rajagopal:1992qz,Grossi:2021gqi}.
Current hydrodynamic simulations do not include the iso-axial charge,  the
pions, or the order parameter as explicit hydrodynamic variables. By
including these variables as explicit degrees of freedom we hope to
increase the  predictive power of hydrodynamic simulations in the
crossover region.

In fact, there is an excess of soft pions relative to the predictions of current ordinary hydrodynamic simulations~\cite{Devetak:2019lsk,Guillen:2020nul,Nijs:2020roc}. We have previously suggested that this excess may reflect the cavalier treatment of chiral symmetry breaking and the $O(4)$ transition in almost all hydrodynamic simulations of heavy ion collisions to date~\cite{Grossi:2021gqi}. To corroborate this suggestion,  we will need to simulate the real-time dynamics of the $O(4)$ phase transition for an expanding fluid. As discussed in \cite{Grossi:2021gqi},  the dynamics of the order parameter matches smoothly onto a pion-hydro effective field theory (EFT) for $T \lesssim T_c$; this pion EFT subsequently  matches onto a kinetic description for soft pion particles coupled to the background fluid flow~\cite{Grossi:2020ezz};  and finally, the kinetics can be used to propagate pions to freezeout with definite predictions for soft pions yields and their correlations.  In addition there is an experimental proposal to measure soft pions and their correlations over a wide range in rapidity~\cite{ALICE}, which is ideally suited to unravel this physics and to probe the (in)applicability of ordinary hydrodynamics in this regime.

As a first step,  in this paper we will  compute the  real-time correlation functions of an $O(4)$ critical system and study their scaling properties. There is considerable theoretical interest in the critical correlators themselves. Many years ago Rajagopal and Wilczek determined that the dynamic universality class of QCD is similar to  ``Model G" of \cite{Hohenberg:1977ym}, where the order parameter $\phi_a = (\sigma, \vec{\pi})$ is not conserved, but has a non-trivial Poisson bracket with vector and axial vector charges~\cite{Rajagopal:1992qz}. They also determined dynamical critical exponent to be $\zeta = \frac{d}{2}$, which we find in the simulations presented here. Because the critical theory must transition between ordinary hydro at high temperatures and a non-abelian superfluid hydro at low temperatures, the expected structure of the hydrodynamic correlations functions is rich~\cite{Grossi:2021gqi}. It would be nice to see this structure in a simulation.

Earlier numerical studies on the critical dynamics of field theories (including $O(4)$ symmetric ones) have been
performed in the
``classical-statistical" framework~\cite{Berges:2009jz,Schlichting:2019tbr,Schweitzer:2020noq}. Given some relativistic quantum field theory, the
high-temperature spectral functions are saturated by their classical
counterparts close to the critical point. Since the non-anomalous symmetries and conservation laws of the classical field theory are
shared with the quantum one,  the classical dynamics belongs to
the same dynamic universality class as the full quantum theory. Of particular relevance to our work was the study done in~\cite{Schlichting:2019tbr}, which studied a classical relativistic $O(4)$ model,
and determined the spectral functions of the order parameter. The spectral functions were
shown to display the appropriate behavior as a function of temperature, and pion
quasiparticle poles were observed in the broken phase. 
Because the classical model has $O(4)$ Noether charge densities, $n_{ab} \sim \phi_{[a} \partial_t \phi_{b]}$, which have a non-trivial
Poisson bracket with the order parameter, the dynamics of this 
model should lie in the universality class of ``Model G".
However, within this set-up 
studying the ``Model G" dynamics is difficult, since  rapidly oscillating UV modes (which  build up the charge densities microscopically)  must be carefully evolved.
Consequently,  a conclusive extraction of the dynamical critical exponent was
not possible, and the interplay between the order parameter and the axial charge was not studied.  
Very recently~\cite{Schweitzer:2021iqk},  the same group adopted an approach for ``Model B" and ``Model D",  which  is somewhat closer in spirit 
to the one taken here for ``Model G", where the charge densities are treated as additional slow variables. In their recent work,  Israel-Stewart-like diffusion models belonging to the specified dynamical class (``B" or ``D")  were simulated, and a careful study of the dynamical scaling function and of their momentum dependence was performed. In particular, this work constitutes an important stepping stone towards the study of ``Model H", which is believed to describe the universality class of the speculated QCD critical point \cite{Son:2004iv}.

An outline of the paper is as follows. In \Sect{sec:model} we discuss the model equations we will solve. Of particular interest is the numerical strategy presented in \Sect{sec:modelnumerics}, which may be useful for other model systems. In
\Sect{sec:thermo} we will review the thermodynamics of the model and fix the non-universal (thermodynamic) parameters of the model. Finally in \Sect{sec:dynamics} we turn to the dynamical properties of the model presenting the
principal results. In \Sect{sec:overview} we present a qualitative overview of the phase transition, and examine the dynamics in the chirally restored limit.
Then in \Sect{sec:broken},  we examine the low temperature limit where the $O(4)$ dynamics should match with the pion EFT. We examine the Gell-Mann-Oakes-Renner relation, and the dissipative pion dynamics proposed by Son and Stephanov~\cite{Son:2001ff,Son:2002ci}.  In the last section, \Sect{sec:dynamicexp}, we examine the scaling of correlation functions along the critical line. We extract the dynamical critical exponent $\zeta$ and find $\zeta \simeq 1.47\pm 0.01\,({\rm stat.})$, which is very close to the predictions of Rajagopal and Wilczek of $\zeta = d/2$. Finally, a short outlook is presented in \Sect{sec:discussion}.



\section{Model}
\label{sec:model}

\subsection{Model equations}
\label{sec:modelequations}

QCD with two degenerate massless quarks is well known to have a second order
phase transition and is in the universality class of the $O(4)$-critical point.
Dynamical properties of a theory near a continuous phase transition are also
universal, but theories with the same static properties can lead to different
dynamical universality classes. Different dynamics arise because of the
existence  or non-existence of conserved charges in the theory~\cite{Hohenberg:1977ym}. As pioneered in
\cite{Rajagopal:1992qz} (see \cite{Grossi:2021gqi} for a recent review)
the dynamics of the QCD $O(4)$-critical point is
the one of an $O(4)$ antiferromagnet, ``Model G'' of \cite{Hohenberg:1977ym}.

Model G consists of an $O(4)$ order parameter\footnote{ Here  $a$ and $b$ denote $O(4)$ indices;  $s$, $s_1$, $s_2$, etc. denote the isospin indices, i.e. the components of $\vec{\pi}$; finally,  spatial indices are notated $i$, $j$ and $k$. The dot product indicates an appropriate contraction of indices when clear from context, e.g. $\phi\cdot \phi=\phi_a \phi_a$,  $\vec{\pi}\cdot\vec{\pi} = \pi_s \pi_s$, and $\nabla\cdot \nabla = \partial_i \partial^i$.}
$\phi_a=(\sigma, \vec{\pi})$ field, and adjoint
charge densities $n_{ab}$. The field $\phi$  is a proxy for the quark condensate
$\llangle \bar q_R q_L \rrangle$, and as a result is not a conserved quantity.
The antisymmetric tensor of charge densities $n_{ab}$ can be decomposed into a vector part,
$n_V^s = \frac{1}{2}\epsilon^{ss_1s_2}n_{s_1s_2}$, and an axial part, $n_A^s =
n_{0s}$.  They represent the original iso-vector $\vec{n}_V  \sim  \bar{q} \gamma^0 \vec{t}_I q  $
and iso-axial $\vec{n}_A  \sim \bar q \gamma^0 \gamma^5 \vec{t}_I q$ charge densities.
The vector current is
exactly conserved for equal quark masses,  while the axial current is
only approximately conserved, since the finite quark mass explicitly breaks the
chiral $SU_{L}(2) \times SU_{R}(2) \sim O(4)$ symmetry.
The equilibrium action (or effective Hamiltonian)
in the presence of an external magnetic field
$H_a=(H,\vec 0)$ parametrizing the explicit symmetry breaking,  takes a Landau-Ginzburg form
\st
\mathcal H \equiv  \int d^3x \left[ \frac{n^2}{4 \chi_0}    +\frac{1}{2} \partial_i \phi_a \partial^i \phi_a + V(\phi) - H \cdot \phi  \right]\, .
\stp
Here $n^2 = n_{ab} n_{ab}$ and
\st
V(\phi) = \frac{1}{2} m_0^2 \, \phi^2  + \frac{\lambda}{4} (\phi\cdot\phi)^2 \, ,
\stp
with $m_0^2$ negative.
The relevant equations of motion for these fields are
\begin{subequations}
\begin{align}
   \partial_t \phi_a  + g_0\,\mu_{ab} \phi_b  &= -\Gamma_0 \frac{\delta \mathcal H}{\delta \phi_a} + \theta_a \, , \label{eq:eom1}\\
   &=\Gamma_0\nabla^2 \phi_a - \Gamma_0 (m_0^2 + \lambda \phi^2)\phi_a + \Gamma_0 H_a + \theta_a\, ,\\
   \partial_t n_{ab}  + g_0 \,\nabla \cdot (\nabla \phi_{[a} \phi_{b]}) + H_{[a}  \phi_{b]} &= \sigma_0 \nabla^2\frac{ \delta \mathcal H }{\delta n_{ab} }    +  \partial_{i} \Xi_{ab}^i  \, , \label{eq:eom2}\\
   &= D_0 \nabla^2 n_{ab} +  \partial_{i} \Xi_{ab}^i  \, .
\end{align}
\end{subequations}

%
Here, for example,   $H_{[a}\phi_{b]}$ denotes the anti-symmetrization, $H_a \phi_b - H_b \phi_a$. $\chi_0$ is the iso-vector and the iso-axial-vector charge susceptibility;  these susceptibilities are equal and approximately constant near the critical point. $\mu_{ab}$ is the chemical potential, $n_{ab}/\chi_0$. The coefficients  $\Gamma_0$ and $\sigma_0$  are the bare kinetic coefficients associated to the order parameter and the charges. The bare diffusion coefficient of the charges is $D_0 = \sigma_0/\chi_0$. The constant $g_0$ is a coupling of the field $\phi$, and has the units of $({\rm action})^{-1}$ in our conventions.
Finally, $\theta_a$ and $\Xi_{ab}$ are the appropriate  noises, which  are defined through their two-point correlations~\cite{Hohenberg:1977ym}
\begin{subequations}
\begin{align}
   \langle \theta_a(t,x)\theta_b(t',x') \rangle &= 2 T_c\Gamma_0 \, \delta_{ab} \, \delta(t-t')\delta^3(x-x') \, ,\label{eq:langevin_var}\\
   \langle \Xi^i_{ab}(t,x)\Xi^j_{cd}(t',x') \rangle &= 2  T_c\sigma_0 \, \delta^{ij} \left(\delta_{ac} \delta_{bd} - \delta_{ad} \delta_{bc} \right) \, \delta(t-t')\delta^3(x-x') \ . \label{eq:langevin_var_cons}
\end{align}
\end{subequations}

The equations of motion
naturally break up into an ideal hydrodynamic  evolution (the left hand side of the equations) with viscous damping (the right hand side of the equations).  If the right hand side is set to zero, it is easy to show that the ideal evolution
leaves ${\mathcal H} = {\rm const}$.   More generally one can write down the
Fokker-Plank equation associated with the stochastic process and straightforwardly show
that the equilibrium probability distribution is
\st
\label{eq:static_distribution}
      P(\phi, n) =  Z e^{-\mathcal H[\phi, n]/T_c }\, .
\stp

 The thermodynamics of this model is recalled in \cite{Grossi:2021gqi} and we will determine some static properties of relevance in Sec.~\ref{sec:thermo}. The real-time correlation functions we will study here are
 \begin{subequations}
 \begin{align}
    G_{\sigma\sigma}(t, k) &\equiv \frac{1}{V} \langle \sigma (t,{\bf k}) \sigma(0,- {\bf k})\rangle_c  \, ,  \\
    G_{\pi\pi}(t, k) &\equiv  \frac{1}{3 V}  \sum_s \langle \pi_s (t,{\bf k}) \pi_s(0,-{\bf k})\rangle_c \, , \\
                            G_{AA}(t, k) &\equiv \frac{1}{3V} \sum_s \langle n_A^s (t,{\bf k}) n_A^s(0,-{\bf k})\rangle_c   \, ,
 \end{align}
 \end{subequations}
 where $\langle \dots \rangle_c$ refers to a connected two point function.
 We will limit this study to $k{=}0$.

 Close to the critical point, the dynamics is expected to be controlled by ``dynamic scaling"~\cite{Hohenberg:1977ym}. In particular,  we expect the time development of our two-point functions  to scale with
 the correlation length $\xi$ as
 \begin{subequations}
 \begin{align}
     G_{\sigma\sigma}(t, k) &= \chi_{\parallel}( k) \, Y_\sigma\left(\Omega\xi^{-\zeta}\, t, \xi k,  z \right) \,, \label{eq:dyn_scal_sigma}\\
     G_{\pi\pi}(t, k) &=  \chi_{\perp}( k) \, Y_\pi\left(\Omega\xi^{-\zeta}\, t, \xi k, z \right) \,,\label{eq:dyn_scal_pi}\\
     G_{AA}(t, k) &=  \chi_{0} \, Y_A\left(\Omega \xi^{-\zeta} \, t, \xi k, z \right) \, . \label{eq:dyn_scal_A}
 \end{align}
 \end{subequations}
 Here the
 functions $\chi_{\parallel}( k), \chi_{\perp}( k)$  are the static order parameter susceptibilities, and depend on $k$ and $\xi$;
 $\chi_0$ is corresponding charge susceptibility which lacks these dependencies;
 $\Omega$ is a  non-universal
 constant normalizing the time; finally,  $z$ is the familiar static scaling variable involving the reduced
 temperature and magnetic field (see below).
 $Y_\sigma$, $Y_\pi$, and $Y_A$ are universal  dynamical scaling functions and $\zeta$ is
 the corresponding dynamical critical  exponent of the theory.
 The expected dynamical critical exponent for ``Model G" is $\zeta=d/2$~\cite{Rajagopal:1992qz}.
 The scaling form (with $\zeta=d/2$) implies that if the correlation length increases by a factor of two, then the characteristic relaxation time increases by a factor of $2^{3/2}$, thereby exhibiting  a ``critical slowing down".

\subsection{Lattice units and matching the model to QCD }
\label{sec:units}

To simulate the model, we begin by  taking $g_0$ and  $T_c$ as our microscopic units
of $({\rm action})^{-1}$ and energy, respectively,   setting $g_0=T_c=1$ in the computer code.
Similarly, we will choose a microscopic length $a$ as the cutoff in our problem, setting the lattice spacing to unity in the code.

As a result of these choices, the quantities we measure directly from our simulations are expressed in lattice units, and they are dimensionless numbers. To convert these quantities to physical predictions, we need to assign a physical value to $g_0$, $T_c$, and $a$.  The critical temperature $T_c$ can  be matched directly to the QCD critical temperature. Once $T_c$ is fixed, $g_0T_c$ is  adjusted so that the model reproduces the pole frequency of the pion.
Lastly, the cutoff $a$ can be adjusted so that our system reproduces the correlation length
of QCD. The aim of this section is to explain this procedure in greater detail.
Before doing so, let us note that our set of units, $g_0  = T_c = a = 1$,
will be implicit both in the figures and the text.
However in this section, and  if necessary for clarity, we  will adopt
a ``hat" notation for variables in lattice units, e.g. $\hat{n}= n \,a^3$
and $\hatchi = T_c \chi \, a^3$ are the dimensionless  charge density and charge susceptibility, respectively.

The model has an $O(4)$ critical point at a critical mass parameter $\hat m_{c}^2(\lambda)$.
At infinite volume and close to the critical point, the dependence of the model condensate (in units of $\sqrt{T_c/a}$) on the mass parameter and magnetic field takes the conventional scaling form~\cite{Engels:2014bra}
\st
\label{eq:scalingform}
\hat{\bar{\sigma}}
=  h^{1/\delta} f_G(z)  \, ,
\stp
where $\delta$ is the critical exponent, and $f_G(z)$ is a universal function with $f_G(0)=1$.
Here $h$ is the reduced magnetic field and  $z$ is the
scaling variable,
\st
\label{eq:scalingfieldsdef}
h \equiv \frac{\hat H}{\hat H_0} \, ,  \qquad   z \equiv \bar t_r h^{-1/\beta \delta}, \quad \mbox{with} \qquad \bar t_r \equiv \frac{\hat m_0^2 -\hat m_c^2}{ \mm^2 }  \ , 
\stp
while $\hat m_c^2(\lambda)$, $\hat H_0(\lambda)$,  and $\mm^2(\lambda)$ are order one non-universal constants that are fit to our
numerical data on thermodynamics (see \Sect{sec:thermo} and \Eq{eq:mc2}). In
physical units
\st
\hat{\bar{\sigma}}  = \frac{\sigma}{B^{O(4)}}\, , \quad \mbox{with} \quad B^{O(4)}= \sqrt{ \frac{T_c}{a}} \, .
\stp

In QCD, the chiral condensate close to the critical point takes the same form ${\llangle \bar q q \rrangle}/{B^\QCD} =  h^{1/\delta} f_G(z)$, but with scaling variables
\st
z \equiv  \left (\frac{T - T_c}{T_c}\right ) h^{-1/\beta\delta} \, , \qquad  h \equiv \frac{m_q c^2}{H_0^\QCD} \ .
\stp
Evidently, to match the two systems we are to equate the scaling variables, $h$ and $z$, and equate the order parameters:
\st
\frac{\llangle \bar{q} q \rrangle}{ B^\QCD }  =\hat{\bar \sigma} = \frac{\bar{\sigma}}{B^{O(4)}}   \, .
\stp

For $H$ small and $T<T_c$,  the universal function $f_G(z)$ behaves as $z^{\beta}$,
and the model condensate
takes the form:
\begin{align}
   \hat{\bar{\sigma} } = \left( \frac{\hat m_c^2 - \hat m_0^2}{ \mm^2} \right)^\beta    \, ,
\end{align}
while chiral condensate takes an analogous form at the corresponding $z$
\begin{align}
   \frac{\llangle \bar q q \rrangle}{B^\QCD} =&  \left( \frac{T - T_c}{T_c} \right)^{\beta}
   \label{eq:scaling_qqbar} \, ,
\end{align}
providing an explicit map between $(T-T_c)/T_c$ and the mass parameter of the model.

The constant $B^\QCD$ has units of (meters)$^{-3}$ and $H^\QCD$ has unit of energy. They can be chosen arbitrarily, but not independently, as the parameter
\begin{align}
  \xi_1^{QCD} = \left(\frac{H_0^\QCD B_0^\QCD}{T_c}\right)^{-1/d} \, ,
\end{align}
fixes a microscopic unit of length. The diverging correlation length of QCD near the  critical point is a universal function times  this length~\cite{onuki_2002}.
As we show in \App{sec:unitsapp},
by choosing
\begin{subequations}
   \label{eq:unitsmatch}
\begin{align}
   a =&  \hat H_0^{1/d} \xi_1^\QCD \,, \\
   g_0 =& \frac{1}{\hbar}  \, ,
\end{align}
\end{subequations}
the model will reproduce both the correlation length  and pole frequency of the pion in QCD.

Having  set three of  our parameters to unity to fix our units of space, time, and energy, we are still left with three more dimensionless parameters which
must be specified, namely
\begin{align}
   \hat\chi_0  \equiv  T_c\chi_0 a^3 \, ,
   \qquad \hat\Gamma_0  \equiv  \Gamma_0  \left(\frac{1}{g_0 T_c a^2} \right) \, , \quad \mbox{and} \quad
     D_0/\Gamma_0 \, .
\end{align}
The susceptibility $\chi_0$ sets the magnitude of charge fluctuations relative to the fluctuations
of the order parameter, while $\Gamma_0$ and $D_0$ determine the relaxation of the order parameter and the charge diffusion, respectively.

Switching to the conventional $\hbar = c=1$ units for this paragraph,
for the system under study there really is only one scale $T_c \sim \Lambda_{\QCD}$.  We expect that the microscopic (i.e. cutoff) length and time are both of order $\sim 1/T_c$. The susceptibility in units of $T_c$ is also of order unity.
Indeed, we expect that all dimensionless constants  are of order unity,
and therefore, in this study  we will take
\st
\hat \chi_0 = 5, \qquad \hat \Gamma_0=1,  \quad \mbox{and} \quad D_0/\Gamma_0 = \frac{1}{3},
\stp
for definiteness.
It may be worthwhile to explore the dependencies on these parameters further, but we have not done so here.

\subsection{Numerical strategy}
\label{sec:modelnumerics}

To simulate the real-time dynamics, we will discretize the stochastic evolution equations in \eqref{eq:eom1} and \eqref{eq:eom2}, placing the system on a spatial lattice of size $L$ and volume $V=L^3$. We briefly present our algorithm in this section; the interested reader can find detailed explanations in App.~\ref{sec:langevinAlgo}.

As the equations naturally separate into an ideal and a dissipative part, we use an ``operator splitting" approach. In spirit, we first evolve  our fields for a short time, neglecting the dissipative part
\begin{subequations}
\label{eq:idealequation}
\begin{align}
   \partial_t \phi_a  &\approx - \mu_{ab} \phi_b \, ,\\
   \partial_t n_{ab}  &\approx -\partial_i (\partial^i\phi_{[a} \phi_{b]}) - H_{[a}  \phi_{b]} \, ,
\end{align}
\end{subequations}
where spatial derivatives are discretized appropriately.
We then neglect the ideal part and solve for the dissipative dynamics
\begin{subequations}
\begin{align}
   \partial_t  \phi_a    &\approx -\Gamma_0 \frac{\partial \mathcal H}{\partial \phi_a} + \theta_a   \, ,\\
   \partial_t  n_{ab}  &\approx \sigma_0 \nabla^2\frac{ \partial \mathcal H }{\partial n_{ab} }    +  \partial_{i} \Xi_{ab}^i \, .
\end{align}
\end{subequations}

Decoupling the equations in such a way allows us to use methods specifically tailored to the two different dynamics. In particular, we use a symplectic integrator to evolve the ideal part, preserving in this way the underlying Poisson bracket structure. To simulate the dissipative Langevin dynamics, we use a Metropolis algorithm. A similar strategy to simulate the Langevin dynamics was used previously to calculate the sphaleron transition rate in hot non-abelian plasmas\footnote{In the sphaleron case the time scales between the metropolis and Langevin times must be carefully matched. In the current simulations, which are near the critical point of the model, this matching  is unnecessary, as the lattice units  and bare parameters  are always adjusted to reproduce  the pion pole frequency and width -- see \Sect{sec:units}.}~\cite{Moore:1998zk}.  
At every lattice site $x$, the order parameter is updated as
\st
\phi_a(t+ \Delta t,x) = \phi_a(t,x) +  \Delta \phi_a \, ,
\stp
where for each flavor index $a$ the increment is
\[
\Delta \phi_a =\sqrt{ 2 \Delta t \Gamma_0 } \, \xi_0 \, .
\]
Here  $\xi_0$ is a random number  with unit variance $\llangle \xi^2_0 \rrangle =1$.
 The update proposal is
accepted with probability ${\rm min} (1, e^{-\Delta \mathcal H})$, where $\Delta\mathcal H$ is the change in the discretized Hamiltonian. If the proposal is rejected, then $\phi(t+\Delta t,x) = \phi(t, x)$.
For small $\Delta \phi_a$
\st
\Delta \mathcal H \approx  \left. \frac{ \partial \mathcal H}{\partial \phi_a} \right|_{\phi_a(x,t)} \Delta \phi_a \, ,
\stp
which can be used to show straightforwardly that the mean and variance of the accepted proposals reproduce the dissipative and stochastic terms of the Langevin process (see also App.~\ref{sec:langevinAlgo} for more details)
\begin{subequations}
\begin{align}
   \overline{\phi_a(t+ \Delta t,x)-\phi_a(t,x)}  &\approx - \Delta t\,  \Gamma_0  \frac{\partial \mathcal H}{\partial \phi_a} \, , \\
      \overline{\left(\phi_a(t+ \Delta t,x)-\phi_a(t,x)\right)^2}  &\approx 2 \Gamma_0 \Delta t  \ .
\end{align}
\end{subequations}
The charges are updated in a similar way, with the extra difficulty that the noise term generated by the updates must be a total divergence. This is tackled by
updating the lattice cells in pairs, making a Metropolis proposal for the charge transfer between two cells -- see App.~\ref{sec:langevinAlgo}.

Using a Metropolis update to solve for the non-ideal part of the dynamics has several advantages over a direct time evolution of the Langevin process. For instance, it allows us to design a scheme whose equilibrium properties are independent of the time-stepping. It also allowed us to use larger time steps compared to a naive discretization of the equations of motion.
The numerical code is implemented with PETSc and MPI~\cite{petsc-web-page,petsc-efficient}.

\section{Statics}
\label{sec:statics}

\subsection{Thermodynamics}
\label{sec:thermo}

Our goal in this section is to fix the static non-universal parameters
of the model from its thermodynamics. 
The magnetization of the system is  an average
over the volume at a given time moment:
\st
M_a(t) \equiv \frac{1}{V} \sum_{x} \phi_a(t,x) \, ,
\stp
and its time average, denoted with $\llangle \ldots \rrangle$,
determines the condensate $\bar{\sigma}$
\st
 \bar{\sigma} \equiv \llangle M_0 \rrangle \, .
\stp
At infinite volume, the dependence of the condensate on the temperature and magnetic field takes the scaling form given in
\eqref{eq:scalingform}.
The non-universal constants $m_c^2$, $H_0$, $\mm^2$ are  fit to our
numerical data on $\bar\sigma$.
We first determine $m_c^2$, then we simulate on the critical line with $m_0^2=m^2_c$ to determine $H_0$, and finally, we simulate at $H=0$ to find $\mm^2$.
Anticipating the results of this section, we obtain with $\lambda =4$
\st
\label{eq:mc2}
m_c^2 = -4.8110(4),  \qquad H_0=5.15(5), \quad
  \mbox{and} \quad \frac{{\mathfrak m}^2}{|m_c^2|} = 1.03(2) \, .
\stp

Following standard technique~\cite{Arnold:2001ir}, we determined the critical coupling of the model $m_c^2$ by measuring Binder cumulants and determining when they cross  a nominal value, which was taken from previous simulations~\cite{Hasenbusch:2000ph}.  Further details are given in  \App{sec:fixingTc}.

 \begin{figure}
    \centering
    \includegraphics[width=0.49\textwidth]{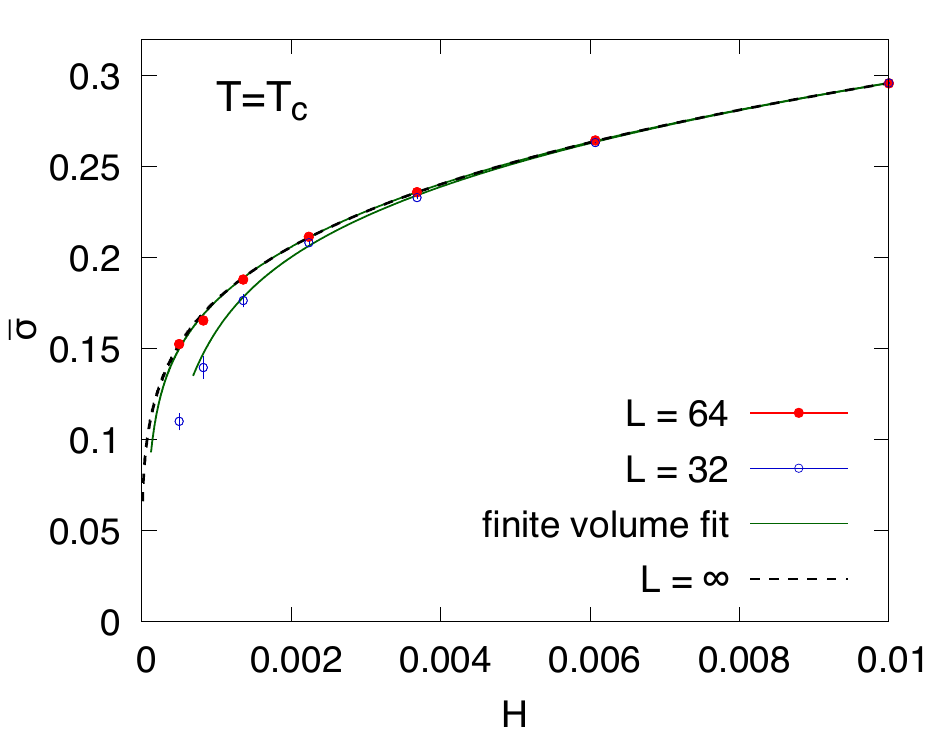}
    \hfill
    \includegraphics[width=0.49\textwidth]{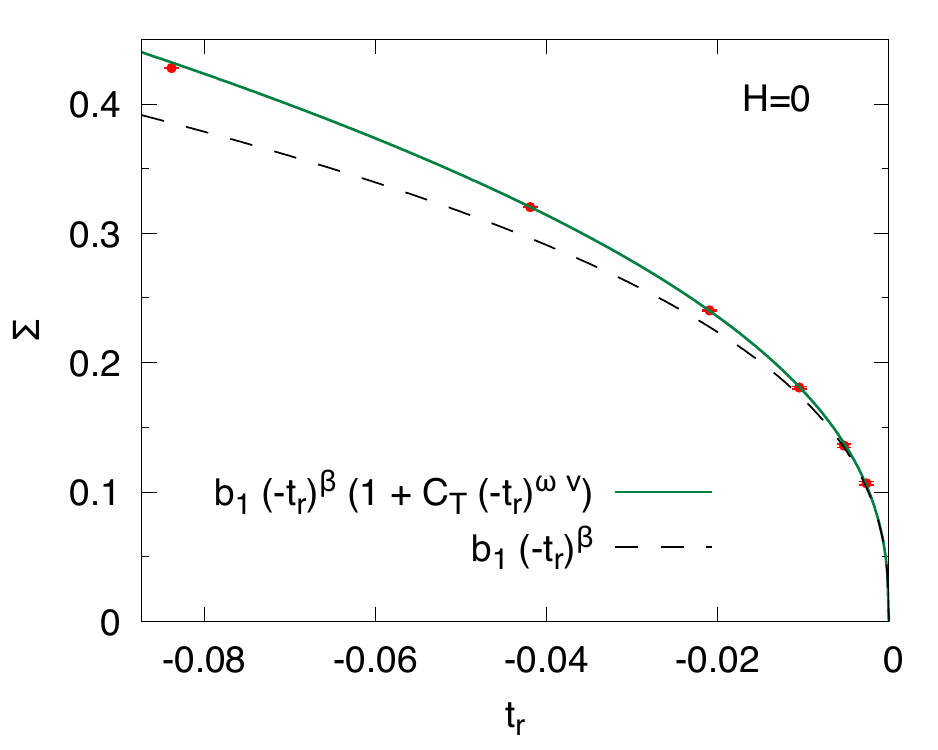}
    \caption{\label{fig:hcrit} {\bf Left:} $\overline{\sigma}$ on the critical line for $L=32$ and $L=64$ together with  a finite volume fit to the data,  which determines the non-universal parameters $H_0$, $L_0$ and $C_H$.  The fit form is taken from Engels and Karsch~\cite{Engels:2014bra} (see text surrounding \Eq{eq:finitevolumefit}). Also shown is the results of the fit at $L=\infty$. {\bf Right:}
       Extracted infinite volume expectation value,
       $\Sigma \equiv \lim_{H\rightarrow 0^+} \lim_{L\rightarrow \infty} \overline \sigma$,   as a
       function $t_r\equiv (m^2_0 - m^2_c)/|m^2_c|$. The fits and extraction procedure are discussed in the text.  Also shown is the fit result without the subleading correction.
 }
\end{figure}

To determine $H_0$ we made a scan on the critical line, with details presented in App.~\ref{sec:thermoh}.
The data for $\bar{\sigma}$ on $32^3$ and $64^3$ lattices on the critical line are shown in the left panel of \Fig{fig:hcrit}. They were fit
to a
a finite size functional form given by Engels and Karsch~\cite{Engels:2014bra}, which fixes the value of $H_0$ given in \eqref{eq:mc2}, and
quantifies finite size corrections. The fit is reasonable and has $\chi^2/{\rm dof}{=}2$.
The magnetization at infinite volume from the results of this fit are shown by the dashed line. We see that already at $L=64$ we are essentially at infinite volume for the range of $H$ considered in this work.
Our dynamical simulations in \Sect{sec:dynamics} are all done with $L^3 = 80^3$. This analysis on the critical line suggests that finite volume corrections are modest.

In the next step we performed simulations at $H=0$  with  $T < T_c$, in order
to fix the non-universal constant ${\mathfrak m}^2$. Details are
 presented in \App{sec:ThermoT}.
The infinite volume magnetization $\Sigma$ at zero field is
defined as
\st
\label{eq:SigmaDef}
\Sigma \equiv \lim_{H\rightarrow 0^+}\lim_{L\rightarrow \infty} \bar{\sigma}\, .
\stp
Extracting the magnetization $\Sigma$ is difficult as, in any finite volume,
\st
 \lim_{H\rightarrow 0}  \left. \bar{\sigma} \right|_{L\,{\rm fixed}}  = 0 \, .
\stp
This is because when $H \Sigma V \sim 1$,
the orientation of  magnetization vector $M_{a}$ begins to wander on
the group manifold, averaging to zero in the limit of zero external
magnetic field.
One way to extract  $\Sigma$ is to look
at the fluctuations of $M_a$,  evaluating $\llangle M^2 \rrangle =\llangle M_a M_a \rrangle$, which is approximately $\Sigma^2$ at large volume.
The leading deviation of $\llangle M^2 \rrangle$ and $\Sigma^2$ at finite volume comes from the fluctuations of long wavelength Goldstone modes, and can be neatly analyzed with a Euclidean pion EFT~\cite{Hasenfratz:1989pk}.
We detail these corrections in \App{sec:ThermoT}, which were essential
to a reliable extraction of $\Sigma(T)$.

Our results for $\Sigma(T)$  are shown in the right panel of \Fig{fig:hcrit},
and are fit with the functional form
\st
\label{eq:sigmavst}
\Sigma =  b_1 (-t_r)^{\beta} \left(1 +  (-t_r)^{\omega \nu} C_{T} \right) \, .
\stp
with critical exponents $\beta$ and $\delta$ from \cite{Engels:2014bra} and $\omega$ from \cite{Hasenbusch:2000ph}.
Here we are using 
\st
t_r \equiv \frac{ m_0^2 - m_c^2}{ |m_c|^2 } \ ,
\stp
instead of $\bar t_r$, and we defined $b_1\equiv(|m_c^2|/{\mathfrak m}^2)^\beta$.  The second term in \eqref{eq:sigmavst} captures the first subleading correction to scaling.

Our fit to $\Sigma(T)$ is shown in the right panel of \Fig{fig:hcrit} and yields
$b_1 = 0.544(4)$ and $C_T =0.20(2)$  with a $\chi^2/{\rm dof}=1.4$.
We have excluded the largest value of $(-t_r)$ from the fit.
For comparison,
we also show the fit results for the first term $b_1(-t_r)^\beta$. Clearly,
for precision work the subleading corrections are important in the temperature
range we are considering.
The parameter $b_1$ determines the scale  ${\mathfrak m}^2$ described
earlier (i.e. ${\mathfrak m}^2 =|m_c^2|\, b_1^{-1/\beta}$) yielding
the results presented in \eqref{eq:mc2}.

To summarize, in this section we have established the non-universal parameters
$m_c^2$ , $H_0$,  and  ${\mathfrak m}^2$  which determines the map between the model
and the conventionally parameterized $O(4)$ critical point.  The results are given in \eqref{eq:mc2}.

\subsection{The Static Pion EFT and Gell-Mann-Oakes-Renner}
\label{sec:stateft}

Before  turning to the dynamics  we will determine the validity of the Euclidean pion EFT discussed above, relegating all details to \App{sec:thermoapppioneft}.
At all temperatures, $O(4)$ symmetry guarantees that the transverse susceptibility is determined by the condensate $\bar\sigma$
\st
\chi_\perp=  \lim_{k\rightarrow 0} G_{\pi\pi}(k) = \frac{\bar{\sigma}}{H} \, ,
\stp
where $G_{\pi\pi}(k)$ is the static correlation function.
At low temperatures the magnitude of the condensate $\sqrt{\phi^2}$  is approximately frozen  to $\bar\sigma$, and the long wavelength order parameter fluctuations are determined  by the  fluctuations in the phase $\varphi$, $\pi_s(x) \simeq \bar\sigma \varphi_s(x)$.
The static action for the Gaussian effective theory describing the phase fluctuations takes the form~\cite{Engels:2009tv,Son:2001ff}
\st
S_{E} = \int {\rm d}^3x \, \frac{1}{2} f^2 \, \nabla \vec{\varphi} \cdot \nabla \vec{\varphi} + \frac{1}{2} f^2 m^2 {\vec{\varphi}}^2 \, ,
\stp
 and makes a definite prediction for the static correlator
\st
G_{\pi\pi}(k) = \frac{\bar\sigma^2}{f^2} \, \frac{1}{k^2 + m^2 }    \, ,
\stp
where $f^2$ is the decay constant and $m$ is the screening mass.
Comparing the predicted  correlator to the susceptibility yields  the Gell-Mann-Oakes-Renner (GOR) relation
\st
f^2 m^2 = H \bar{\sigma} \, .
\stp
At a finite negative $z$, the GOR relation is only approximate, receiving corrections due to fluctuations of the $\sigma$ field.
We have fit the static $\pi\pi$ correlator to find the decay constant $f^2$ and the screening mass $m^2$ at a nominal point in the broken phase, $z=-2.2011$ and $H=0.003$ (see \Fig{fig:chiplot}). Comparing  $f^2m^2$ to $H\bar\sigma$
yields
\begin{align}
   \label{eq:GOR1}
   \frac{f^2 m^2}{H\bar\sigma} =& 1.006 \pm 0.007\,({\rm stat}) \, .
\end{align}
Evidently, already at $z=-2.2$,  the Euclidean pion EFT
works to better than a percent.  

Having studied the statics of the pions, in the next section we will  turn to the dynamics, making use of these results in \Sect{sec:broken}.

\section{Dynamics}
\label{sec:dynamics}
\subsection{The simulations}

\begin{figure}
   \centering
  \includegraphics[width=0.6\textwidth]{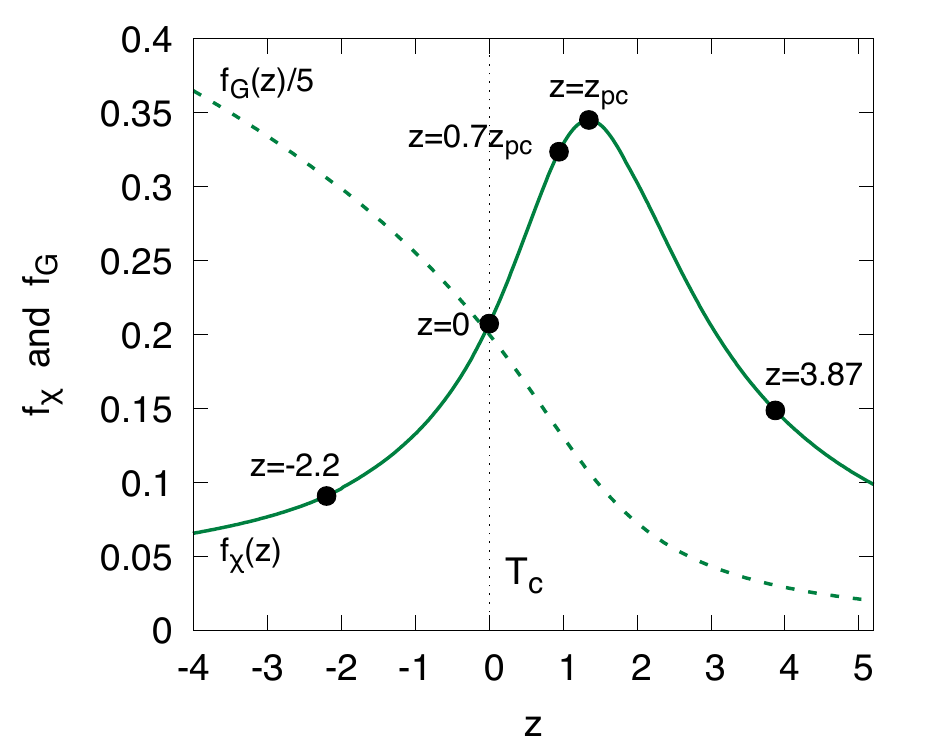}
  \caption{ \label{fig:chiplot} A parametrization of the longitudinal susceptibility $\chi_{\parallel} \propto f_{\chi}(z)$ and mean magnetization $\bar{\sigma} \propto f_{G}(z)$ taken from the simulations of Engels and Karsch~\cite{Engels:2014bra}. The black points are the $z$ values which will be simulated in this work.  Further simulation details are given in \Fig{fig:runs}. }
\end{figure}

To get an overview of the phase diagram, in \Fig{fig:chiplot} we show the scaling function of the magnetization $f_G(z)$ and the corresponding
function for longitudinal susceptibility  $f_\chi$~\cite{Engels:2014bra}
\st
\chi_{\parallel} = \frac{\partial \bar\sigma}{\partial H} = \frac{h^{1/\delta -1}}{H_0} f_\chi(z) \, .
\stp
\begin{figure}
  \centering
  \begin{tikzpicture}
    \footnotesize
    \draw (-7,0) -- (7,0);

    \node at (-6,2){$H=0.003$};
    \draw[] (-6,0) -- (-6,0.3);
    \node at (-6,-0.5){$z=-2.2011$};
    \node at (-6,-1.0){$m_0=-4.99128$};

    \draw[dashed,thick] (-2.5,0) -- (-2.5,1.2);
    \node at (-2.5,3.5){$H=0.01\phantom{0}$};
    \node at (-2.5,3){$H=0.006$};
    \node at (-2.5,2.5){$H=0.004$};
    \node at (-2.5,2){$H=0.003$};
    \node at (-2.5,1.5){$H=0.002$};

    \node at (-2.5,-0.5){$z=0$};
    \node at (-2.5,-1.0){$m_c=-4.81100$};
    \node[right] at (-2.5,0.6){${\large T_c}$};

    \node at (0.1,2){$H=0.003$};
    \draw[] (0.1,0) -- (0.1,0.3);
    \node at (0.1,-0.5){$z=0.94429$};
    \node at (0.1,-1.0){$m_0=-4.73366$};

    \node at (2.7,2){$H=0.003$};
    \draw[] (2.7,0) -- (2.7,0.3);
    \node at (2.7,-0.5){$z_{\rm pc}=1.34899$};
    \node at (2.7,-1.0){$m_0=-4.70052$};

    \node at (6,2){$H=0.003$};
    \draw[] (6,0) -- (6,0.3);
    \node at (6,-0.5){$z=3.86978$};
    \node at (6,-1.0){$m_0=4.49406$};

    \node[blue] at (-6, 3.2){{\color{blue}\bf Broken phase}};
    \node[red] at (3, 3.2){{\color{MyLightRed}\bf Unbroken phase}};
  \end{tikzpicture}
  \caption{\label{fig:runs}Overview of the different simulations used in this works. In all cases,  $\chi_0 = 5$, $\Gamma_0 = 1$, $D_0=\frac13$. The simulations are run on a lattice of volume  $V=80^3$  for $10^6$ timesteps (see App.~\ref{sec:langevinAlgo} for a discussion of the algorithm and the size of the steps).  The first $10^4$ are discarded to ensure the system has thermalized.}
\end{figure}
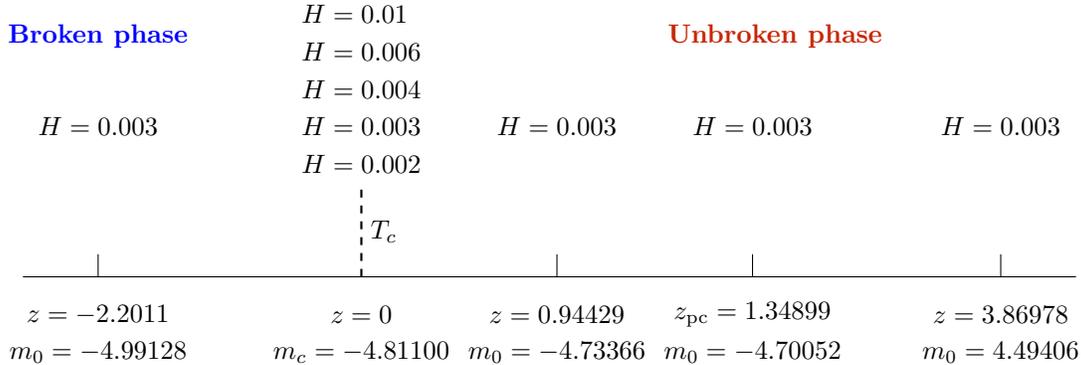

The susceptibility shows a prominent maximum at the pseudocritical point
with $z$ value of  $z_{\rm pc} \simeq 1.35$. In order to
scan the dynamics of the transition, we have
performed real-time simulations at the black points.
We also made a scan on the
critical line $z=0$ for various values of the magnetic field.
The dynamical parameters as well as the run times and other information
is gathered in Fig.~\ref{fig:runs}.

\subsection{Overview}
\label{sec:overview}

We will start by presenting an overview of the critical dynamics as the temperature is scanned across the phase transition.
At high temperatures, the order parameter is small and simply dissipates through the damping term in the equations of motion. Since there is no preferred direction, the longitudinal and transverse order parameters excitations,  $\delta \sigma$ and $\vec{\pi}$,  are nearly degenerate.
In the vector channel, the total charge is constant in time and the dissipation affects only non-zero Fourier modes, which are not studied here.
The situation is different in the axial channel, since the axial charge is not conserved. However, the explicit symmetry breaking term in the action, $H \sigma$, is tiny, since it is  proportional to the magnetic field $H$ (or quark mass)  and the order parameter, which is  small at high temperatures, $\bar \sigma \propto H$.
As a result, the axial charge will dissipate rather slowly over a time scale  of order $H \bar{\sigma}/\chi_0 \propto H^2$. In this regime, the dynamics of the axial charge is unrelated to the pions.
However, as we lower the temperature, the order parameter acquires an $H$-independent expectation value, and the axial channel gets modified;  the order parameter field and the axial charge are now entangled.
In the deeply broken phase at low temperatures, the axial charge and the transverse part of the field  will no longer just dissipate. Indeed,  their dynamics become intrinsically locked,  and they acquire the  quasiparticle characteristics of the Goldstone modes associated with  the broken symmetry.
By contrast, in this regime the longitudinal excitation of the order parameter (the $\sigma$) has a large mass and its dynamics remains purely dissipative.

\begin{figure}
   \centering
%
%
%
%
%
%
\includegraphics[width=\textwidth]{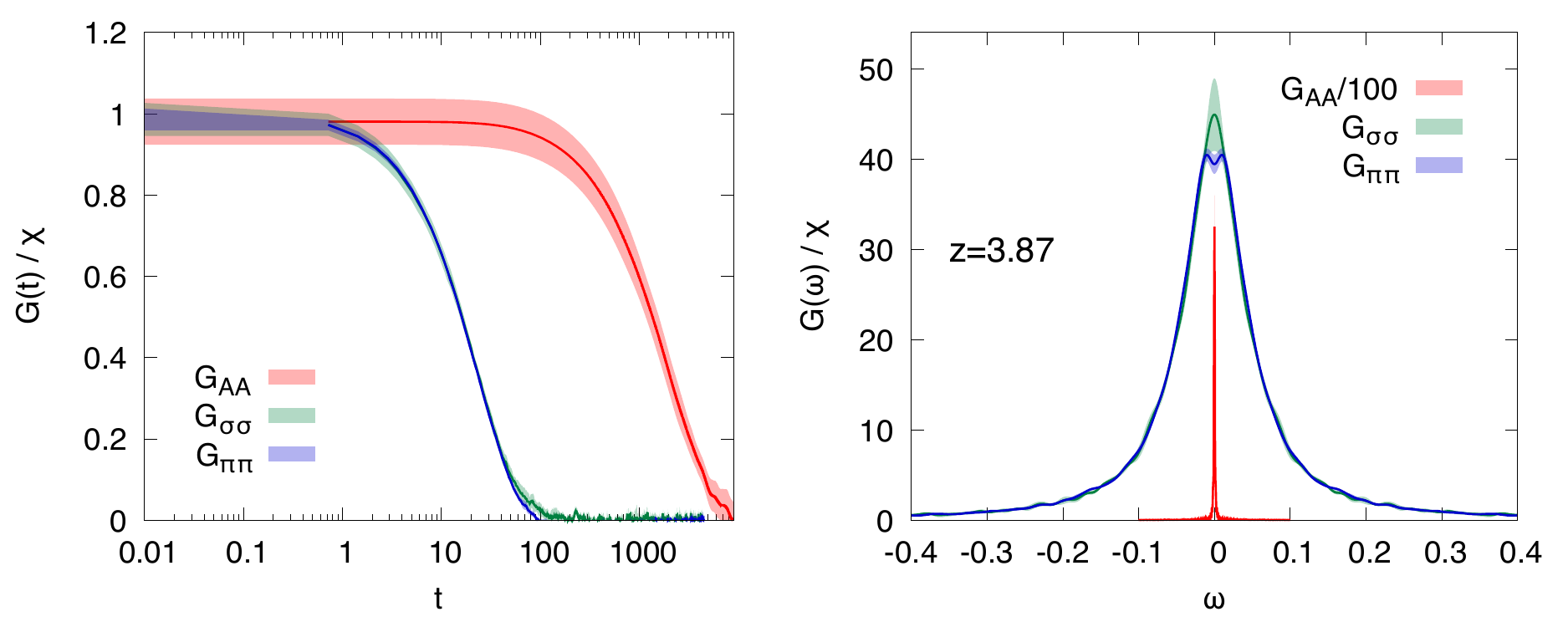}
   \caption{\label{fig:unbroken} \textbf{Left:} Statistical correlation functions  in the unbroken phase as a function of time for  the axial charge,  $\sigma$, and $\pi$.  In this regime the $\sigma$ and $\pi$ channels become degenerate, and the axial charge is almost conserved. The logarithmic $x$-axis emphasizes that the relaxation time of the axial charge is orders of magnitude longer than the one of the  $\sigma$ and $\pi$. \textbf{Right:} The corresponding correlators
      as a function of frequency.
   }
\end{figure}

These qualitative behaviors are precisely observed in our data. In Fig.~\ref{fig:unbroken}, we start by showing  the results of a simulation performed in the unbroken phase, $z=3.87$. In the left plot we show the statistical correlator  for the $\sigma$, $\pi$, and axial channels as a function of time. Noting that the $x$-axis is on a logarithmic scale, the slow dissipation in the axial channel is apparent. It is also apparent that the $\sigma$ and $\pi$ channels are almost degenerate and dissipate on a much shorter timescale. This is also clearly seen in the corresponding Fourier transforms (right), where $\sigma$ and $\pi$ correlators appear as a single dissipative peak, which is much broader than corresponding peak in the axial correlator. 

\begin{figure}
   \centering
%
%
%
\includegraphics[width=\textwidth]{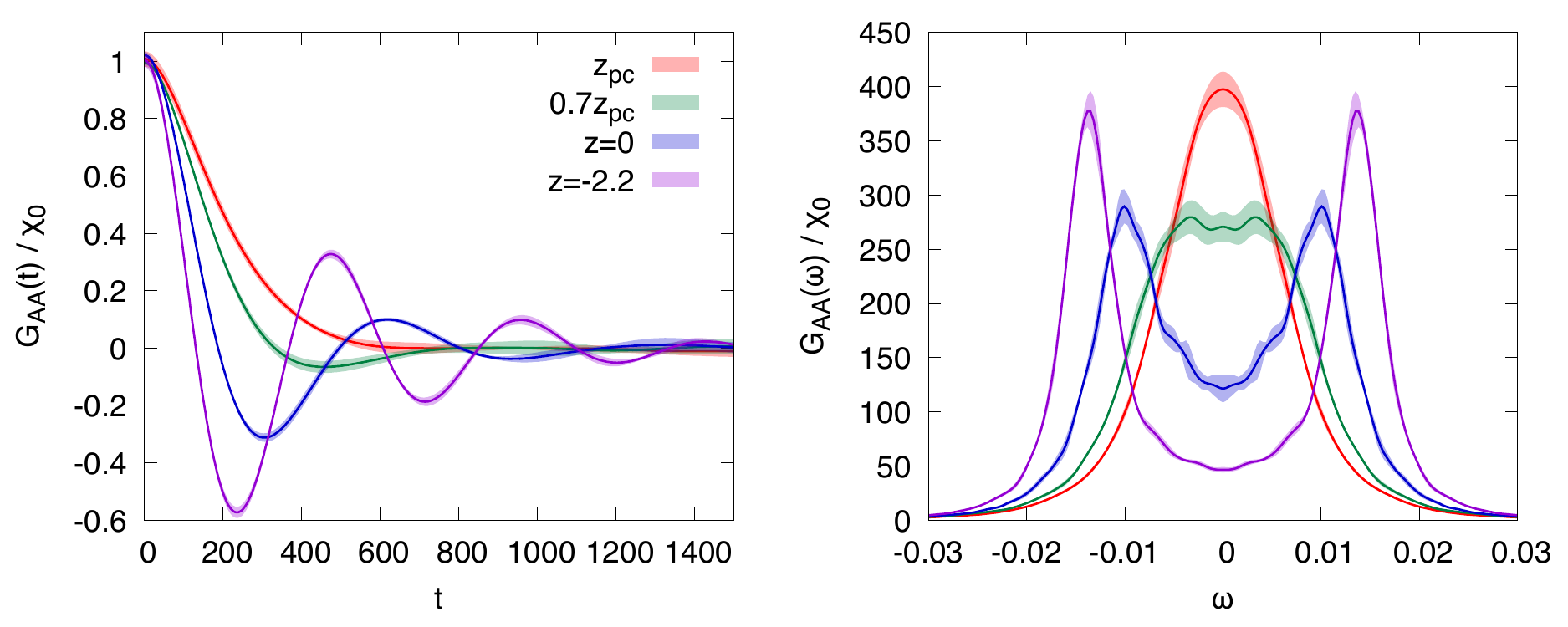}
   \caption{\textbf{Left:} The axial charge correlator as a function of time with $z$ spanning the phase transition. \textbf{Right:} The corresponding correlators in the frequency domain.\label{fig:pseudocritical}}
\end{figure}

In Fig.~\ref{fig:pseudocritical}, we show the behavior of the axial charge
correlator at the pseudocritical and critical temperatures, and at  lower
temperatures, in the broken phase. In the left panel, we show the correlation
functions as a function of time, while in the right panel we show their Fourier
transforms. At the pseudocritical temperature (the red curves),  the axial charge correlator is still purely dissipative,  but the peak is much broader than in Fig.~\ref{fig:unbroken}, indicating that the charge is  no longer  approximately conserved.
As we lower the temperature to $T_c$ (the blue curves), we start seeing the emergence
of propagating pions, which appear as oscillations in the
correlator as a function of time, or equivalently, as quasi-particle peaks  in the Fourier transform. At
the critical temperature  there are no drastic changes (this is expected in
a finite magnetic field), and the correlator behaves as it does in the broken phase, with
propagating pions which are clearly visible in the axial channel. As one moves
further into the broken phase (the purple curves), the pion peaks become
increasingly separated, and the real-time pion EFT 
discussed below 
becomes valid
(see Sect.~\ref{sec:broken}).

It seems that around the pseudocritical temperature $z_{\rm pc}$ (the red curve) the axial charge propagator starts changing its behavior from purely dissipative to quasiparticle-like. Indeed,  at $z=0.7z_{\rm pc}$ (the green curve), i.e. slightly below the pseudocritical point, the dissipative peak is already quite deformed, which reflects the nascent formation of the two quasiparticle peaks. 

\begin{figure}
   \centering
   \includegraphics[width=\textwidth]{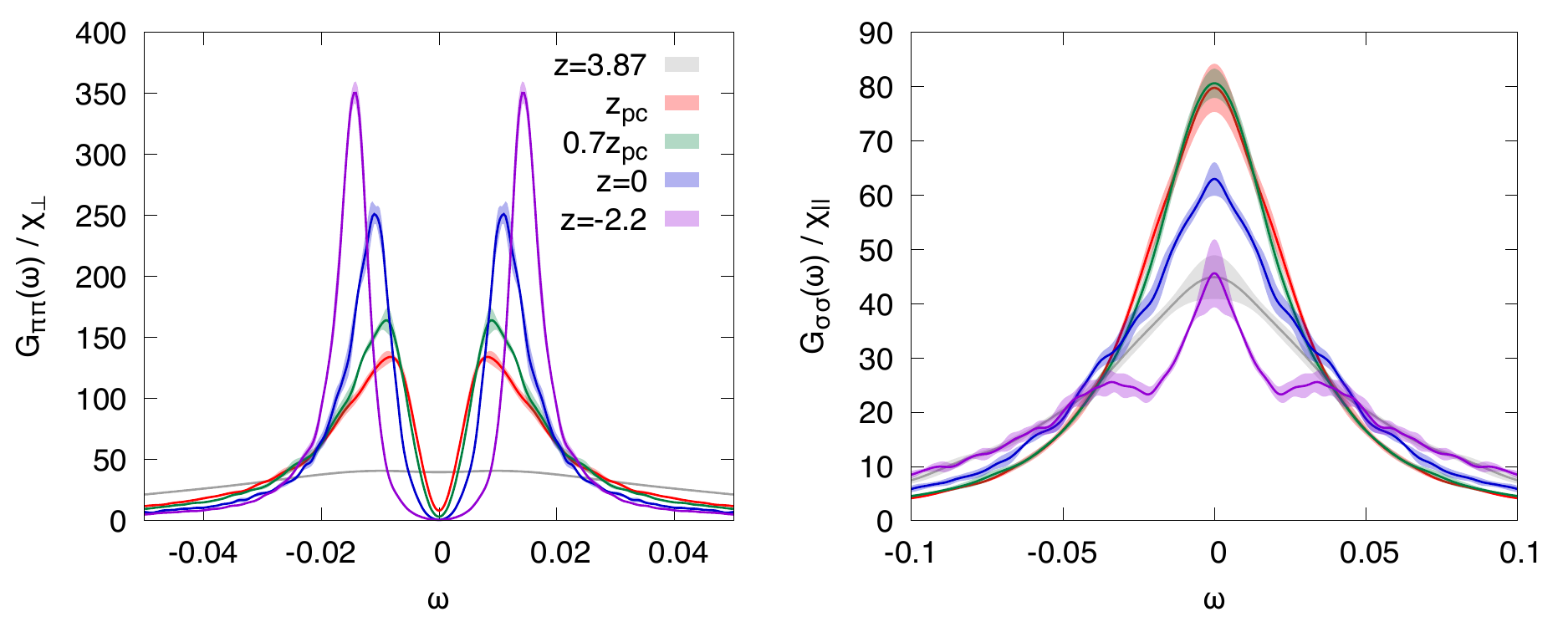}
   \caption{\label{fig:broken}  \textbf{Left:} The Fourier transform of the  $\pi\pi$ correlator with $z$ spanning the phase transition. \textbf{Right:}
   The corresponding  $\sigma\sigma$ correlators. Note the different scales used in the two panels. In particular, the gray curves in the left and right panels are almost identical, as shown in Fig.~\ref{fig:unbroken}.}
\end{figure}

In the left and right panels of Fig.~\ref{fig:broken},  we show the corresponding statistical correlators  for the $\pi$ and $\sigma$ fields as a functions of frequency, with $z$ spanning the phase transition.
In the deeply unbroken phase the two channels are mostly indistinguishable (the grey bands), as pointed out before.
Lowering the temperature to the pseudocritical point, the pseudoscalar channel acquires a double peak structure, while the scalar channel remains purely dissipative.
Going further down in temperature, the quasi-particle peaks in the pseudoscalar channel separate.
Interestingly at $z_{\rm pc}$, the pion correlator already has a quasiparticle peak,  while the axial charge correlator is still dissipative (\Fig{fig:pseudocritical}); only past the pseudocritical point do their correlation functions become closely related.

\subsection{Broken phase: pion EFT}
\label{sec:broken}

Deep in the broken phase, the fluctuations of the order parameter
are dominated  by the phase fluctuations $\pi_s(t,x) \simeq \bar\sigma \varphi_s(t,x)$, which are tightly correlated to the axial charge fluctuations through the Josephson constraint, $\partial_t \vec{\varphi}  \simeq \vec{\mu}_A$. The
dissipative hydrodynamic theory for the phase fluctuations has been worked out in \cite{Son:1999pa,Son:2001ff,Grossi:2020ezz},
and provides a real time analog of the static Gaussian effective theory described in \Sect{sec:stateft}.


The linear response of hydrodynamic theory has been analyzed in \cite{Son:2001ff,Grossi:2021gqi},
and  the hydrodynamic prediction
for the dynamical correlators in the ${\bm k}=0$ case is
\begin{align}
  G_{\pi \pi}(\omega) &= \frac{2 \chi_\perp \Gamma m^2 \omega^2}{(-\omega^2 + m_p^2)^2 +\omega^2(\Gamma m^2)^2}\, , \label{eq:piEFTphiphiCor}\\
  G_{AA}(\omega) &= \frac{2 \chi_0 \Gamma m^2 m_p^2}{(-\omega^2 + m_p^2)^2 +\omega^2(\Gamma m^2)^2} \ . \label{eq:piEFTAACor}
\end{align}
Here $m_p^2$ is the pole mass of the pion excitation, $m$ is the transverse static screening mass, $\Gamma$ is a dissipative coefficient correcting the Josephson constraint,  and finally $\chi_0$ and $\chi_\perp$ are the appropriate static susceptibilities, which are required to normalize these expressions
\begin{align}
   \int \frac{\mathrm{d} \omega}{2\pi} G_{\pi \pi}(\omega) &= \chi_\perp \, , \\
   \int \frac{\mathrm{d} \omega}{2\pi} G_{AA}(\omega) &= \chi_0 \, .
\end{align}

The fact that the pions are pseudo-Goldstone bosons, and correspondingly that the axial current is partially conserved (PCAC), leads to the well-known and remarkable property that the dynamical pole mass $m_p$ can purely be computed from the static properties discussed in \Sect{sec:stateft}. In particular, at low-enough temperatures, we have a finite temperature Gell-Mann-Oakes-Renner (GOR) relation \cite{Pisarski:1996mt,Son:2001ff,Son:2002ci}
\begin{align}
  m_p^2 = v^2 m^2 = \frac{H\bar\sigma}{\chi_0} \, , \label{eq:OaksRenner}
\end{align}
where  $v^2\equiv{f^2}/{\chi_0}$ is the pion velocity.

Already in Fig.~\ref{fig:pseudocritical} we saw the appearance of pion excitations. We will now try to assess the validity of the pion EFT. To do so, we
attempt to fit expressions \eqref{eq:piEFTphiphiCor}-\eqref{eq:piEFTAACor} from our statistical correlators. To perform these fits, we first fix the normalizations by extracting from our data the susceptibilities, $\chi_0$ and $\chi_\perp$. We then use a two parameter model, involving  $m_p$ and $\Gamma_p = \Gamma m^2$, and simultaneously fit the statistical correlators in the $\pi$ and axial channels.

\begin{figure}
   \centering

  \includegraphics[width=\textwidth]{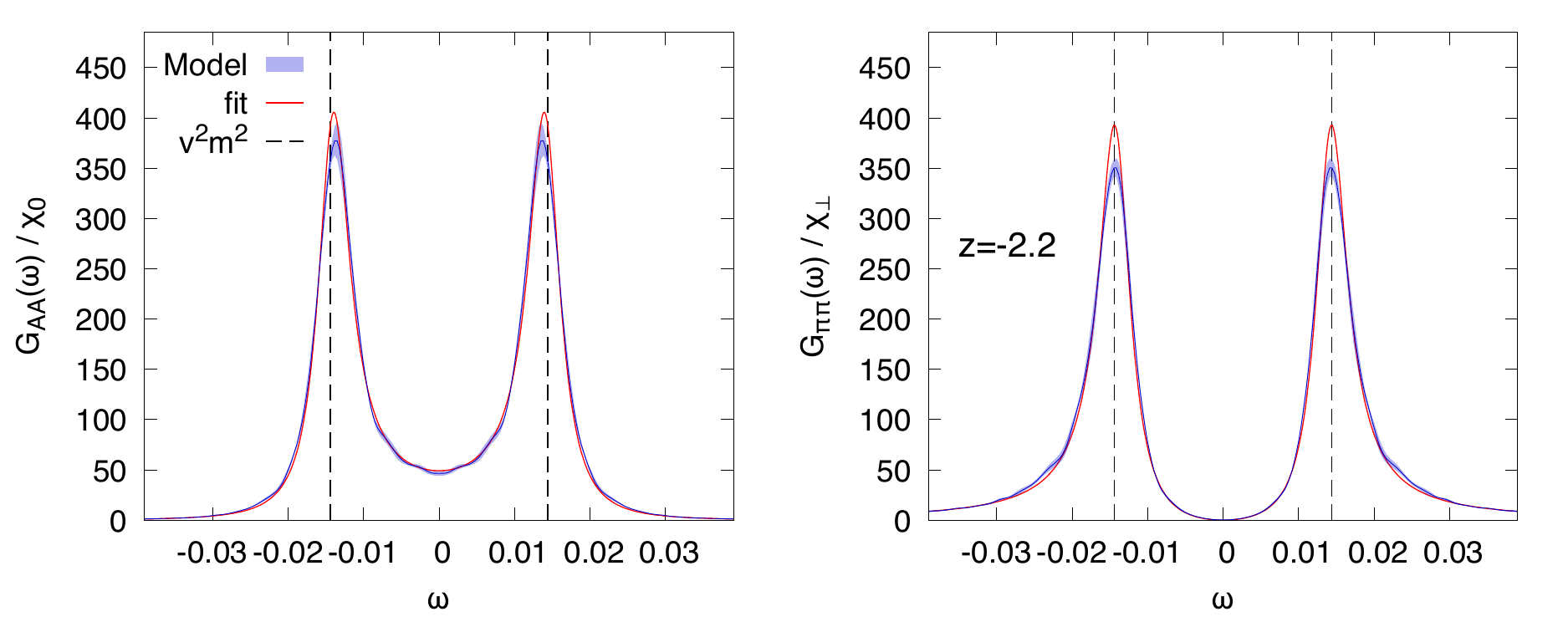}
  \caption{\label{fig:broken_fits}  Statistical correlators  for the $\pi$ and axial channels together with the result of a global fit to the functional form from the chiral hydrodynamic theory (see text). The fitted parameters are $m_p=(1.4387\pm 0.0005\, ({\rm stat.}))\cdot 10^{-2}$ and $\Gamma_p=(5.088\pm 0.005\,({\rm stat.}))\cdot 10^{-3}$, with a $\chi^2/{\rm dof}$  of $1.93$.}
\end{figure}

Results of these fits are shown in Fig.~\ref{fig:broken_fits}, yielding parameters
\begin{align}
   m_p&=(1.4387\pm 0.0005\,({\rm stat.}))\cdot 10^{-2}\,,  \label{eq:res_mp}\\
   \Gamma_p&=(5.088\pm 0.005\,({\rm stat.}))\cdot 10^{-3}\, , \\
  \chi^2/{\rm dof}&=1.93 \, .
\end{align}
Although the width is still pretty large, $\Gamma_p/2m_p \simeq 0.17$,
we find good agreement between the numerical data and the pion EFT, with only small noticeable deviations around the maxima of the two point functions.

This extraction of the pole mass allow us to verify the dynamical part of the GOR relation. Referring the reader again to \Sect{sec:stateft}  for the corresponding extraction of the static quantities, we find
\begin{align}
   \frac{H\bar\sigma}{\chi_0} \cdot \frac{1}{m_p^2}= 1.011 \pm 0.001\,({\rm stat.})\label{eq:res_mpstat} \ .
\end{align}
We see again that already at $z=-2.2$, the deviations from GOR are remarkably small, of order $1\%$, which could be due to corrections of order $\sim (\Gamma_p/2m_p)^2$. Note also that part of this $1\%$ deviation could be due to some remaining systematic errors in our time evolution, see App.~\ref{sec:idealstep} for more details.

\subsection{Critical line: dynamical scaling}
\label{sec:dynamicexp}

Moving on to the critical line $z=0$,  we consider the scaling of the critical dynamics. Focusing on $\vec{k}=0$ modes and recalling that the (e.g. longitudinal) correlation length scales as
\begin{align}
   \xi = \xi_c H^{-\nu_c} \,,
\end{align}
on the critical line, the ``dynamic scaling hypothesis" \eqref{eq:dyn_scal_sigma}-\eqref{eq:dyn_scal_A} gives us the following scaling forms
for the correlators on the critical line
\begin{align}
  \frac{G_{\sigma\sigma}(t,H)}{\chi_{\parallel}} &=  Y_\sigma^c\left( H^{\zeta\nu_c}  t \right)\,, \label{eq:dyn_scal_sigma_k0}\\
  \frac{G_{\pi\pi}(t,H)}{\chi_{\perp}} &=  Y_\pi^c\left(H^{\zeta\nu_c} t \right)\, ,\label{eq:dyn_scal_pi_k0}\\
  \frac{G_{AA}(t,H)}{\chi_{0}} &=  Y_A^c\left(H^{\zeta\nu_c}  t \right)\label{eq:dyn_scal_A_k0} \, ,
\end{align}
with for example, $Y^c_A\left(H^{ \zeta \nu_c}t \right)=Y_A\left(\Omega \, \xi^{-\zeta} t,0,0 \right)$.

\begin{figure}
   \centering
   \includegraphics[width=\textwidth]{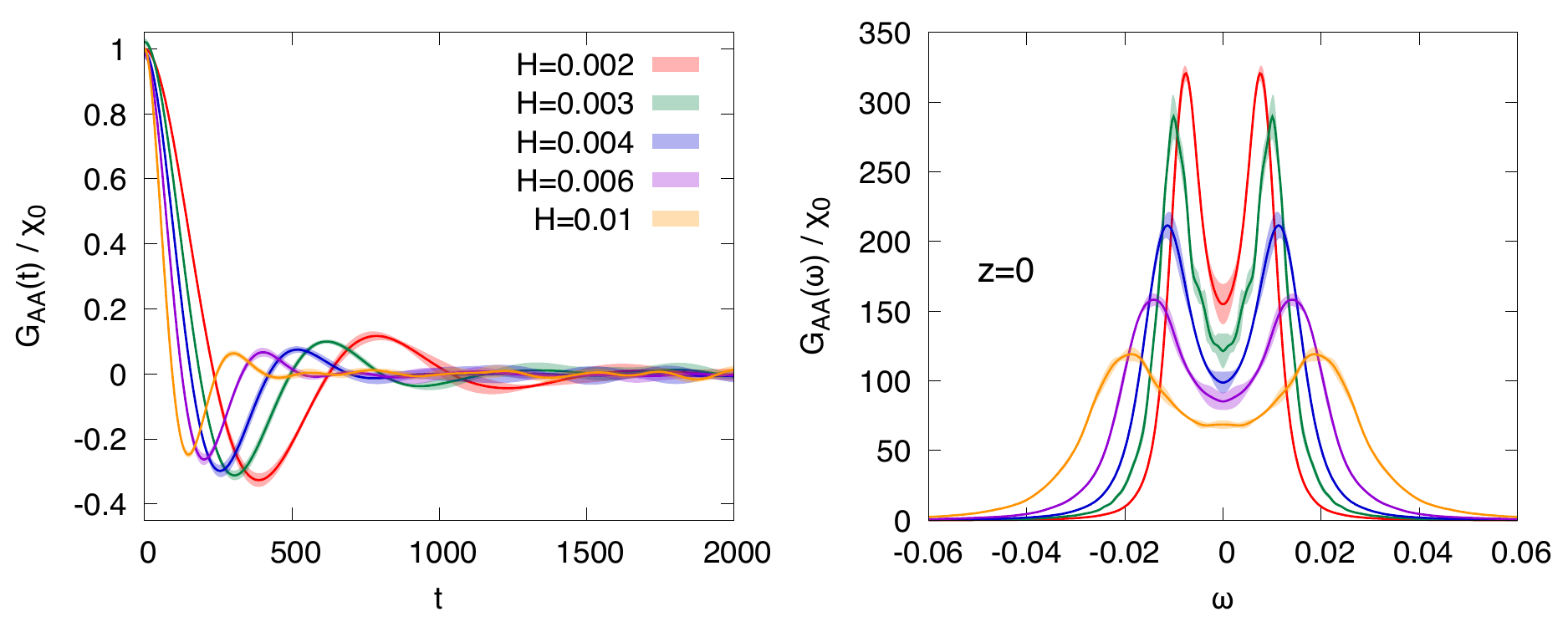}
   \caption{\label{fig:critical1} \textbf{Left:} Time dependent axial charge correlation functions for different magnetic fields on the critical line, $z=0$. \textbf{Right:} Corresponding statistical correlator in frequency space. }
\end{figure}

To verify the validity of the hypothesis and to determine the dynamical exponent $\zeta$, we studied a set of simulations at $m_0^2=m_c^2$ for  $H=0.002, 0.003, 0.004, 0.006, 0.01$. We present the results obtained for the axial-axial channel in Fig.~\ref{fig:critical1}. The left plot shows the time-dependent correlator $G_{AA}(t)$ for the different magnetic fields, while the right panel displays  its corresponding Fourier transform $G_{AA}(\omega)$.
Qualitatively at least the curves show a scaling behavior.
\begin{figure}
   \centering
   \includegraphics[width=\textwidth]{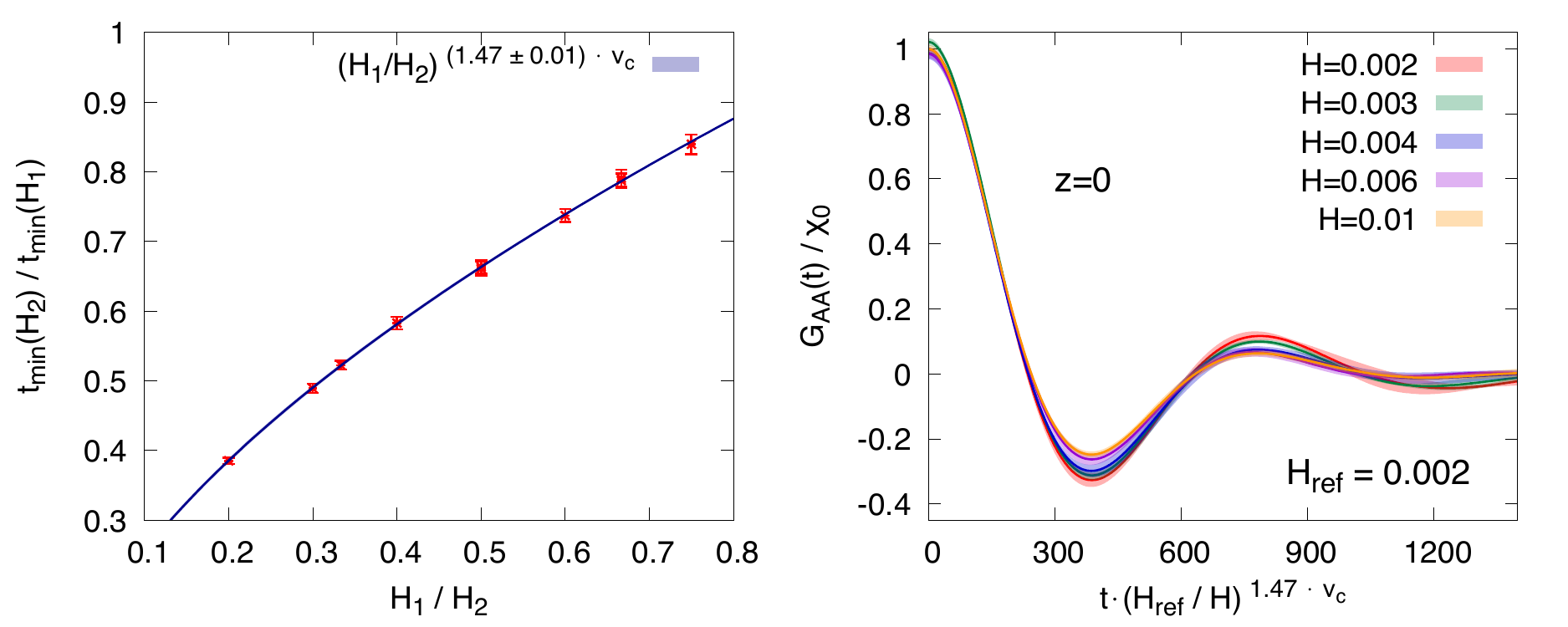}
   \caption{\label{fig:critical2} \textbf{Left:} Ratios,  $t_{\rm min}(H_2)/t_{\rm min}(H_1)$, extracted from the first minima of Fig.~\ref{fig:critical1} as a function of $H_1/H_2$ (see text). On the critical line, we expect this ratio to be described by a universal critical exponent $t_{\rm min}(H_2)/t_{\rm min}(H_1)= (H_1/H_2)^{\zeta\nu_c}$. Our best fit gives $\zeta=1.47\pm 0.01$.  \textbf{Right:} Time dependent axial correlation functions as a function of an appropriately rescaled time variable.}
\end{figure}

To quantitatively assess the scaling ansatz \eqref{eq:dyn_scal_A_k0} and to extract the exponent $\zeta$ from our data,
we located the time when  $G_{AA}(t,H)$ reaches its first minimum, $t_{\rm min}(H)$, which can be determined with reasonable accuracy.
From the scaling ansatz, we see that, given two magnetic fields $H_1, H_2$, we expect
\begin{align}
  \frac{t_{\rm min}(H_2)}{t_{\rm min}(H_1)} = \left (\frac{H_1}{H_2}\right)^{\zeta\nu_c} \ .
\end{align}
We show this ratio as a function of $H_1/H_2$ in the left panel of Fig.~\ref{fig:critical2}.
The data are well described by the power law form,  and
we obtain a nominal value for the dynamical exponent of
\begin{align}
   \zeta_{\rm fit}=1.47\pm 0.01 \, ,
\end{align}
taking $\nu_c=0.4024$ from \cite{Engels:2014bra}.

\begin{figure}
   \centering
   \includegraphics[width=\textwidth]{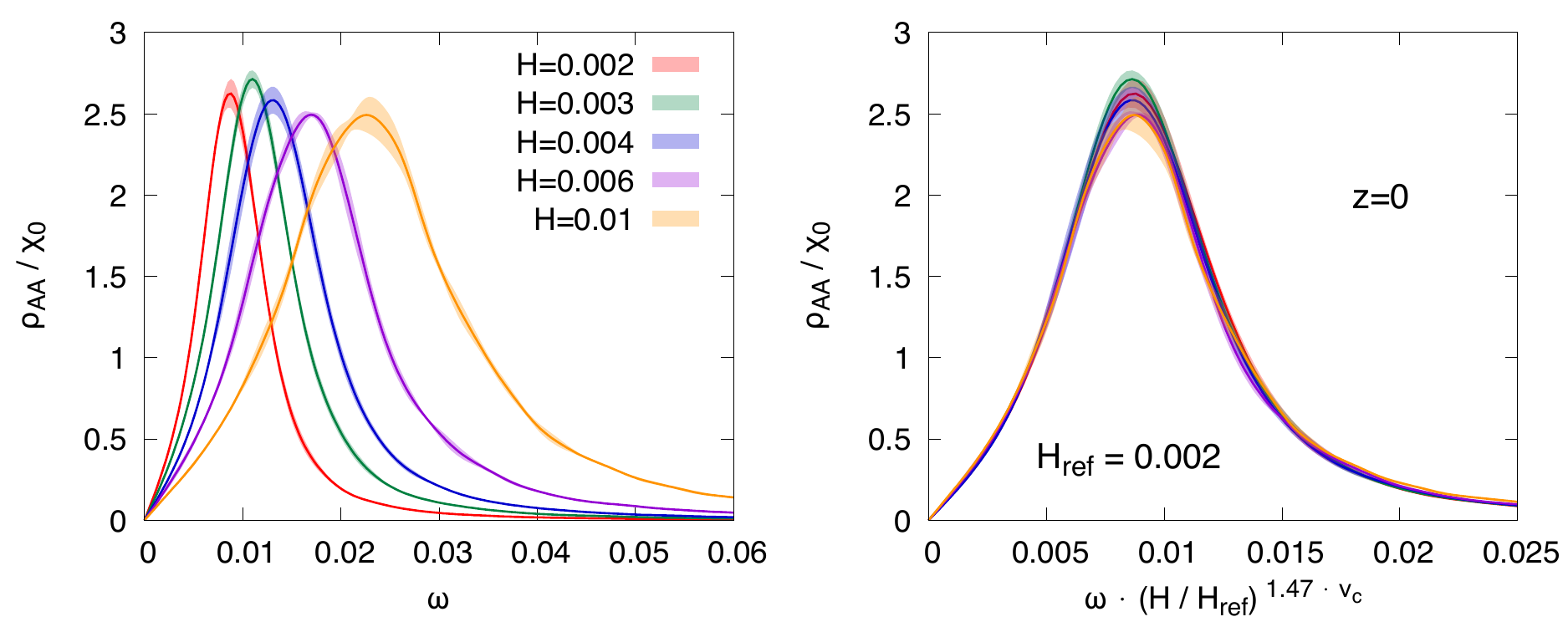}
   \caption{\label{fig:critical3}  \textbf{Left:} Axial charge spectral function on the critical line. \textbf{Right:} Axial spectral function as a function of an appropriately rescaled frequency variable. In both cases, we see the expected data collapse.}
\end{figure}

With an estimate of the critical exponent in hand, we can verify the ansatz \eqref{eq:dyn_scal_A_k0}. Indeed by appropriately rescaling times and frequencies, we expect to see our correlators $G_{AA}(t,H)$ and spectral function
\begin{align}
  \rho_{AA}(\omega,H) = \omega G_{AA}(\omega,H) \, ,
\end{align} collapse to a single curve. The scaling of $G_{AA}(t,H)$ is shown in the right panel of Fig.~\ref{fig:critical2}, while
the scaling $\rho_{AA}(\omega, H)$ is shown in \Fig{fig:critical3}. To obtain this data collapse, we have rescaled the time and inverse frequency by $(H_{\rm ref}/H)^{\zeta_{\rm fit}\nu_c}$ with $H_{\rm ref}=0.002$.

\begin{figure}
   \centering
   \includegraphics[width=\textwidth]{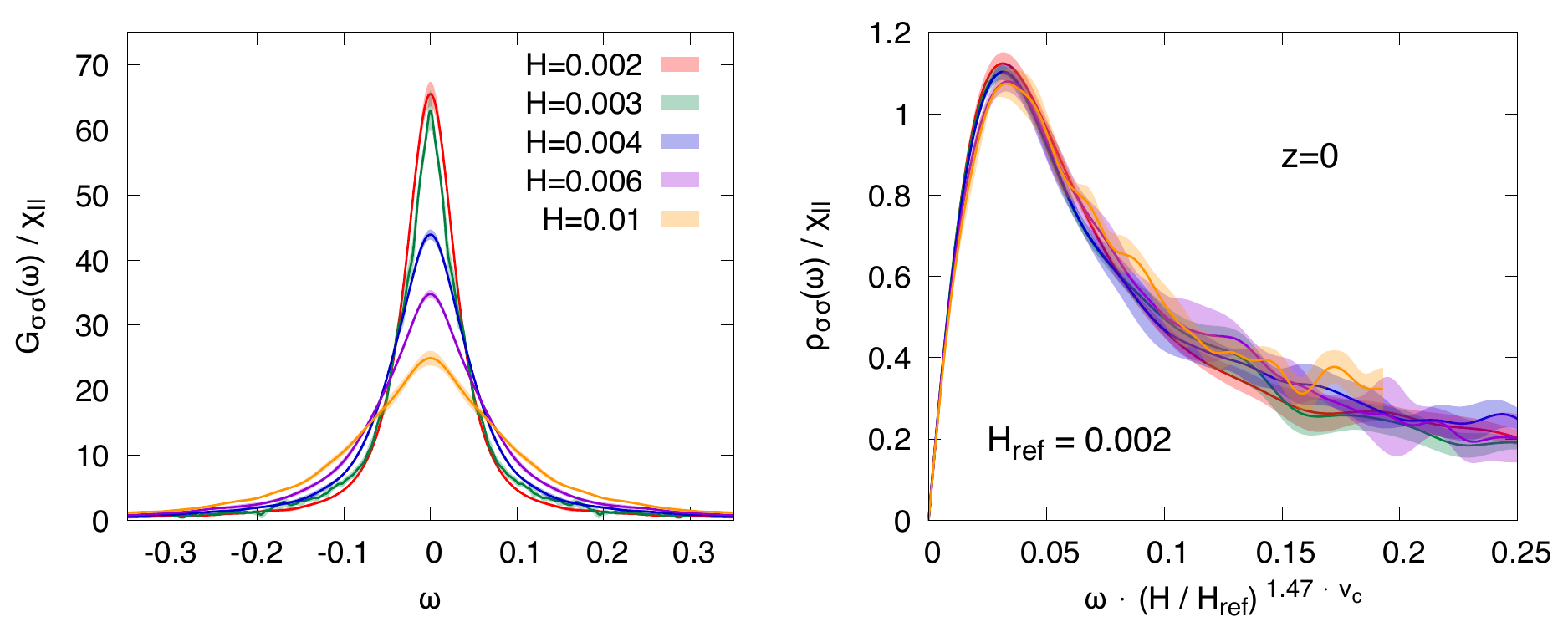}
   \caption{\label{fig:criticaldsigma} \textbf{Left:} Statistical correlator in the $\delta \sigma$ channel, on the critical line. \textbf{Right:} Corresponding spectral function as a function of rescaled frequencies. The estimated critical exponent used for the rescaling is determined from the axial channel. The dynamical scaling hypothesis is satisfied and we observe a collapse of the different curves. }

\end{figure}

Dynamical scaling is also expected to hold in the other channels and in particular we expect the same $\zeta_{\rm fit}$ to  govern the dynamics in the $\sigma$ channel. This indeed happens, which we illustrate in Fig.~\ref{fig:criticaldsigma} by showing the $\sigma\sigma$ correlator (left) and the corresponding collapsed spectral function (right).

Before moving on, let us emphasize that our numerical estimate of the critical dynamical exponent is close to the critical scaling prediction \cite{Rajagopal:1992qz,Hohenberg:1977ym},  $\zeta=d/2$.
Considering, for example, the small violations of scaling seen in \Fig{fig:critical2}, we do not consider the deviation of $\zeta_{\rm fit}$ from $d/2$ to be significant.

\section{Discussion}
\label{sec:discussion}
In this work, we numerically studied the  universal critical dynamics relevant to two-flavor QCD  close to the chiral phase transition. More precisely, we simulated the dynamics of an $O(4)$ antiferromagnet,  ``Model G'' of \cite{Hohenberg:1977ym,Rajagopal:1992qz}. After reviewing the model and explaining our conventions in \Sect{sec:model}, we performed some ``scale setting" in \Sect{sec:statics}, where we studied the thermodynamic properties of the model and extracted  the relevant non-universal constants.
We also determined some of the static properties of pions in the broken phase such as their screening masses and decay constants.

With this data in hand, we moved on to the main section of this work,  Sec.~\ref{sec:dynamics}, which studied the dynamics.
Focusing on correlators at zero spatial momentum, we first performed a scan in temperature across the phase transition. We qualitatively confirmed that the dynamics takes place as expected, by studying the real-time correlation functions in the $\sigma\sigma$, $\pi\pi$ and axial-axial channels. At high temperature, the $\sigma$ and $\pi$ are degenerate and the axial charge is almost conserved. In the broken phase, the $\sigma$ remains purely dissipative,  while the $\pi$ propagates and carries axial charge. In particular, we were able to observe that the coupling of the $\pi$ to the axial charge precisely happens in the vicinity of the pseudocritical point, $z_{\rm pc}$, defined as the line in the phase diagram where the static susceptibility peaks. This observation is yet another link between the static and dynamical properties of this critical model.

We also performed a quantitative study of the pion properties in the broken phase. We were able to fit the dynamical correlator to a particle resonance ansatz predicted by the chiral hydrodynamic effective theory, and extract the pole mass and decay width. Furthermore, we verified that the Gell-Mann-Oakes-Renner relation, which relates the dynamical pole mass of the pions to their static screening mass, holds at the sub-percent level. Last but not least, we performed a set of simulations along the critical line and extracted the dynamical critical exponent $\zeta=1.47\pm0.01\, ({\rm stat})$, very close to the critical scaling prediction $\zeta=1.5$~\cite{Rajagopal:1992qz}.

The numerical determination of $\zeta$ can be considered as a first step towards a complete quantitative characterization of the dynamics of the $O(4)$ antiferromagnet. Such a characterization would include additional studies  at finite spatial momentum as in~\cite{Schweitzer:2021iqk}, and a more complete investigation of the dynamics in the chiral limit at finite volume with an appropriate  real-time EFT.
(The corresponding  finite volume static EFT   was written down long ago~\cite{Hasenfratz:1989pk}, and was helpful in the thermodynamic analysis in \Sect{sec:ThermoT}).
In order to use the model to analyze heavy-ion data  as discussed in \cite{Grossi:2020ezz,Grossi:2021gqi},  it will be important to analyze the  critical $O(4)$ dynamics for an expanding fluid, which introduces a rich hierarchy of scales.
Finally, it will be interesting to apply the algorithm presented in App.~\ref{sec:langevinAlgo} to other stochastic and critical systems.

\begin{acknowledgments}
The authors would like to thank Stony Brook Research Computing and  Cyberinfrastructure , and the Institute for Advanced Computational Science at Stony Brook University for access to the SeaWulf computing system, which was made possible by  a National Science Foundation grant (\#1531492).

In addition, this  research used resources  resources of the National Energy Research Scientific Computing Center (NERSC), a U.S. Department of Energy Office of Science User Facility located at Lawrence Berkeley National Laboratory, operated under Contract No. DE-AC02-05CH11231 using NERSC award 001-ERCAP6306.

   This work is supported  by
the U.S. Department of Energy, Office of Science, Office of Nuclear Physics,
grants Nos.
DE\nobreakdash-FG\nobreakdash-02\nobreakdash-08ER41450. AS is supported by the Austrian Science Fund (FWF), project no. J4406.
E.G. is supported in part by the GLUODYNAMICS project funded by the ``P2IO LabEx (ANR-10-LABX-0038)" in the framework ``Investissements d’Avenir" (ANR-11-IDEX-0003-01) managed by the Agence Nationale de la Recherche (ANR), France.
\end{acknowledgments}

\appendix

\section{The model on the Lattice}
\label{sec:langevinAlgo}

\subsection{Relating lattice units to QCD}
\label{sec:unitsapp}

Continuing the discussion of \Sect{sec:units}, our goal is
to fix the spatial cutoff $a$ in units of meters, so that
the model will reproduce the physical correlation length. In the computer code
the cutoff is the lattice spacing, and is set to unity.
Similarly the times are measured in units of $1/(g_0 T_c)$, and we will
set the $g_0$ in physical units so that the model reproduces 
the pion pole frequency.
The results of this section are summarized by \Eq{eq:unitsmatch}

For any critical system with canonically normalized magnetic Hamiltonian $\Delta \mathcal H =\int d^3x H  \sigma$, and mean order parameter of the form  $\bar\sigma = B h^{1/\delta} f_G(z)$ with $h=H/H_0$, the combination of parameters $H_0 B/T_c$ has dimensions $({\rm length})^{-d}$ and defines a  non-universal length:
 \st
 (\xi_1)^{-d} \equiv \frac{H_0 B}{T_c} \, .
\stp
The longitudinal correlation length of the generic critical system takes the form
\st
\xi =  \xi_1  h^{-\nu_c} \,  f_{\xi}(z) \, ,
\stp
where $f_{\xi}(z)$ is a universal function, including its normalization\footnote{The function $f_{\xi}(z)$ for the longitudinal correlation length is proportional to $\hat g_{\xi}^{L}(z)$ of \cite{Engels:2009tv},
with the proportionality constant  given by universal amplitude ratios.
 With some patience one finds,
 $f_{\xi}(z) = (Q_c R_\chi)^{1/d} (Q_2^{L}/\delta R_\chi)^{\nu/\gamma} {\hat g^{L}}_\xi(z)$.
}.

In the $O(4)$ model
where
\st
H_0^{O(4)}=\sqrt{T_c/a^{d+2} }\, \hat{H}_0, \quad \mbox{and} \quad B^{O(4)}= \sqrt{T_c/a^{(d-2)}} \, ,
\stp
 the length scale $\xi_1$ evaluates to
\st
\xi_1^{O(4)} =  a \, \hat{H_0}^{-1/d}  \, .
\stp
In QCD we have $\xi_1^{\QCD} = (H_0^{\QCD} B^\QCD/T_c)^{-1/d}$, leading
to the identification given in the text
\st
\label{eq:latticea}
a =  \hat H_0^{1/d} \xi_1^\QCD \, .
\stp

Next we discuss the dynamics. There is
a time scale set by the frequency of the pion   pole\footnote{This formula assume that the total charge operators $Q_{ab}$ are unitless and satisfy the $O(4)$ commutation relations
   \st
   [ Q_{ab} , Q_{cd} ] = i \left( \delta_{ac} Q_{bd}  + \delta_{bd} Q_{ac} - \delta_{ad} Q_{bc} -  \delta_{bc} Q_{ad} \right) \, .
   \stp
   The  susceptibility is defined by the averages
   \st
   \llangle Q_{ab} Q_{cd} \rrangle =T\chi_0 V \left(\delta_{ac}\delta_{bd} - \delta_{ad}\delta_{bc} \right) \, .
   \stp
}
\begin{align}
   (m_p^\QCD)^2 \equiv& \frac{1}{\hbar^2} \, \frac{m_q c^2 \llangle \bar{q} q \rrangle }{\chi_0} \, , \\
   =& \frac{H_0^{\QCD}}{\hbar^2} \, \frac{h \llangle \bar{q} q \rrangle }{\chi_0} \, ,
\end{align}
In the $O(4)$ model it is easy that the corresponding frequency in physical
units is
\begin{align}
   m_p^2 =& \frac{g^2_0 H \bar \sigma }{\chi_0}  = g^2_0 \,T_c \hat H_0    \, \left(\frac{h \,  \hat{\bar{\sigma}}/a^3} {\chi_0} \right) \, ,
\end{align}
Comparing the two expressions, using $\llangle \bar q q \rrangle = B^\QCD \hat{\bar{\sigma}}$ and the identification
\st
h = \frac{m_q c^2}{H^\QCD} = \frac{\hat H}{\hat H_0} \, ,
\stp
leads to the result
\st
\label{eq:g02}
g_0^2 =  \frac{1}{\hbar^2}   \, .
\stp
Eqs.~\eqref{eq:latticea} and \eqref{eq:g02} are presented in the body of the text in \Eq{eq:unitsmatch}.

\subsection{Overview of the algorithm}
In this section we will describe the update algorithm in detail.
We first discretize the fields on a spatial lattice (with lattice spacing $a=1$) writing  the effective Hamiltonian as
\st
\label{eq:latticeaction}
\mathcal H =  \sum_{x,e_i} \frac{1}{2}  (\phi(t,x+e_i) - \phi(t, x))^2   + \sum_x  V(\phi(t,x)) - H \sigma(t,x)   + \sum_x
\frac{n^2(t,x) }{4\chi_0} \, .
\stp
Here $x=(x_1,x_2,x_3)$ labels the lattice sites, and $e_i =e_1, e_2, e_3$ is a unit vector in the corresponding direction. Variational derivatives
in the equations of motion get replaced by ordinary derivatives, $\delta {\mathcal H}/\delta \phi \rightarrow \partial {\mathcal H}/\partial \phi$, etc.

The equations of motion for  $U\in [\phi_a, n_{ab}]$ can
be written schematically
\st
\partial_t U=  \mathcal O_A(U) + \mathcal O_B(U) + \mathcal O_C(U) \, .
\stp
The first operator describes the evolution under the ideal equations of motion, while the second
two operators describe the dissipative dynamics of the order parameter $\phi$ and the charges $n_{ab}$ respectively.
We will use operator splitting to solve for the total time evolution. The most straightforward procedure
is to  update the fields  sequentially for a small period time $\Delta t$
\st
U \xrightarrow{A} U \xrightarrow{B} U \xrightarrow{C} U \, .
\stp
More explicitly  we have: \\

\noindent {\bf A Stage:}
\begin{align}
   \partial_t \phi_a  =& - \mu_{ab} \phi_b  \, ,  \\
   \partial_t n_{ab}  =& \partial_i (\phi_a \partial^i\phi_{b} - \phi_b \partial^i \phi_b ) - (H_{a}  \phi_{b} - H_{b} \phi_a )  \, ,
\end{align}
\noindent {\bf B Stage:}
\begin{align}
   \partial_t \phi_a  =& -\Gamma_0 \frac{\partial \mathcal H}{\partial \phi_a} + \theta_a  \, ,\\
   \partial_t n_{ab}  =& 0 \, ,
\end{align}
\noindent {\bf C Stage:}
\begin{align}
   \partial_t \phi_a  =& 0 \, ,  \\
   \partial_t n_{ab} =& \sigma_0 \nabla^2\frac{ \partial \mathcal H }{\partial n_{ab} }    +  \partial_{i} \Xi_{ab}^i \, .
\end{align}
We will view each step as part of a Markov Chain.

A technical complication is that the $C$ step takes approximately six times longer than the $B$ step, because there
are many more random numbers to generate. The ideal step $A$ also is about twice slower than the $B$ step.
So as a practical matter, for a complete step over a time $\Delta t$  we will take the following updates
\st
   ABB\,ABB\, ABB\; C \, ,
\stp
where the time increment for $B$ is $\Delta t_B = \Delta t/6$ while the time step for $A$ is $\Delta t_A=\Delta t/3$.  An optimal time step thermalizes modes of order
the lattice spacing in a short period of wall-time. We have found
$\Delta t = 0.24 /\Gamma_0$ is approximately optimal (see below), for the algorithm discussed here.

\subsubsection{Ideal step}
\label{sec:idealstep}

In order to perform our ideal step, let us first rewrite the ideal part of our continuous equation as follows
\begin{align}
  \partial_t \phi_a &= -\frac{n_{ab}}{\chi_0}\phi_b \, ,  \label{eq:idealPhi} \\
  \partial_t n_{A}^s &=\partial_{i}( \sigma\partial^i\phi_s - \phi_s\partial^i \sigma ) - H \phi_s \label{eq:idealA} \, ,\\
  \partial_t n_{V}^s &= \epsilon^{ss_1s_2}\,\partial_i(\phi_{s_1}\partial^i\phi_{s_2})  \, . \label{eq:idealV}
\end{align}
Eq.~\eqref{eq:idealPhi} makes it apparent that the ideal evolution of the order parameter is simply an $O(4)$ rotation by the currents. More explicitly, in the $O(4)$-algebra matrix notation, we have

\begin{equation}
   \partial_t \phi = -\frac{i}{\chi_0}N \phi \label{eq:idealPhiBis} \, ,
\end{equation}
with
\begin{equation}
  N(t) = \vec n_A(t)\cdot \vec K + \vec n_V(t)\cdot \vec J \, ,
\end{equation}
and $\vec{K}, \vec J$ the generators of $so(4)$
\begin{align}
K_1 = -i \begin{pmatrix}
 0 & 1 & 0 & 0 \\
-1 & 0 & 0 & 0 \\
 0 & 0 & 0 & 0 \\
 0 & 0 & 0 & 0 \\
\end{pmatrix} \ \
K_2 = -i \begin{pmatrix}
 0 & 0 & 1 & 0 \\
 0 & 0 & 0 & 0 \\
-1 & 0 & 0 & 0 \\
 0 & 0 & 0 & 0 \\
\end{pmatrix} \ \
K_3 = -i \begin{pmatrix}
 0 & 0 & 0 & 1 \\
 0 & 0 & 0 & 0 \\
 0 & 0 & 0 & 0 \\
-1 & 0 & 0 & 0 \\
\end{pmatrix} \\
J_1 = -i \begin{pmatrix}
0 & 0 &  0 & 0 \\
0 & 0 &  0 & 0 \\
0 & 0 &  0 & 1 \\
0 & 0 & -1 & 0 \\
\end{pmatrix} \ \
J_2 = -i \begin{pmatrix}
0 & 0 & 0 &  0 \\
0 & 0 & 0 & -1 \\
0 & 0 & 0 &  0 \\
0 & 1 & 0 &  0 \\
\end{pmatrix}\ \
J_3 = -i \begin{pmatrix}
0 &  0 & 0 & 0 \\
0 &  0 & 1 & 0 \\
0 & -1 & 0 & 0 \\
0 &  0 & 0 & 0 \\
\end{pmatrix} \ .
\end{align}
 In particular, \eqref{eq:idealPhiBis} can be solved as
 \begin{equation}
   \phi(t+\delta t) = \exp\left(-\frac{i}{\chi_0}\int_{t}^{t+\delta t} \mathrm{d} t' N(t') \right) \phi (t)\ .
 \end{equation}

With this in mind,  before describing our time evolution, we need to discretize \eqref{eq:idealA} in space. For $f$ and $g$ functions evaluated on a discrete lattice, we discretize terms of the sort $\partial_x\left(g\partial_x f \right)$ in a straightforward way by integrating over finite volume cells, e.g. in one dimension
 \begin{align}\nonumber
   \frac{1}{a}\int_{x_{i-\frac12}}^{x_{i+\frac12}} \mathrm{d}x\, \partial_x\left(g\partial_x f \right) &\simeq   \frac{g|_{x_{i+\frac12}}}{a^2}\left(f|_{x_{i+1}} - f|_{x_{i}} \right) -\frac{g|_{x_{i-\frac12}}}{a^2}\left(f|_{x_{i}} - f|_{x_{i-1}} \right) \\
   &\simeq \frac{1}{2a^2}
   \Big[\left(g|_{x_{i+1}}+g|_{x_{i}}\right)\left(f|_{x_{i+1}} - f|_{x_{i}} \right) \\ \nonumber
  & \qquad   \qquad  \qquad  - \left(g|_{x_{i}}+g|_{x_{i-1}}\right) \left(f|_{x_{i}} - f|_{x_{i-1}} \right)  \Big] ,
 \end{align}
 where we have approximated the value $g|_{x_{i+\frac12}}$ at each interface as the mean
 of the central one, $g|_{x_{i+\frac12}}=\frac12 (g|_{x_{i+1}}+g|_{x_{i} })$.
Using the shorthand notation $f_{\pm i}\equiv f(t, x\pm e_i)$,  leads us to define the following discrete evolution kernels
\begin{align}
   \mathcal{K}_V^{s} &=\frac{\epsilon^{ss_1s_2}}{a^2} \sum_{i=1}^3  \Big(
      \pi_{s_1} \pi_{s_2,+i}  - \pi_{s_1,-i} \pi_{s_2}
\Big) \, , \\
  \mathcal{K}_A^{s}&=-\pi_s H + \frac{1}{a^2} \sum_{i=1}^3  \Big(
     \sigma \pi_{s,+i} - \pi_s \sigma_{+i}   -\sigma_{-i} \pi_{s} + \pi_{s,-i} \sigma
\Big) \ .
\end{align}

To evolve this system, we use a ``position Verlet''-like symplectic integration. We start by computing $\Phi$ at half-integer steps, use it to evolve the currents by a time step $\delta t$ and finish updating $\Phi$ by an extra half-time step, which gives
\begin{align}
   \phi\left(t+\frac12 \delta t\right)& =  \exp\left(-\frac{i}{\chi_0}\frac{\delta t}{2} N(t) \right) \phi(t)\label{eq:discrot1} \, , \\
   n_{A}^s\left(t+\delta t\right) &=n_{A}^s\left(t\right) + \delta t   \mathcal{K}_{A}^{s} \, ,\\
   n_{V}^s\left(t+\delta t\right) &=n_{V}^s\left(t\right) +\delta t   \mathcal{K}_{V}^{s}\, ,\\
  \phi\left(t+\delta t\right) &=  \exp\left(-\frac{i}{\chi_0}\frac{\delta t}{2} N(t+\delta t) \right) \phi\left(t+\frac{\delta t}{2}\right)\ \label{eq:discrot2} .
\end{align}
A practical way to perform the rotations in \eqref{eq:discrot1} and \eqref{eq:discrot2} is to rewrite the $O(4)$ rotation as a direct product of  $SU_{L}(2) \times SU_{R}(2)$ and to use the explicit form of the $SU(2)$ matrices for a given set of angles.

The ideal evolution is associated to the conservation of the
the discretized energy, $\mathcal H$, for $\delta t\rightarrow 0$.
Our symplectic evolution leads to a violation $\Delta E = {\mathcal H}(t+ \delta t)- {\mathcal H}(t)  \sim O(\delta t^2)$.
One approach to this violation would be to just ignore it. Then the equilibrium action will be modified slightly by terms of order $O(\delta t^2)$ from \eqref{eq:latticeaction}, shifting $T_c$ by a small amount. We have seen indications of
these shifts but have not explored this in detail. Instead we have added
a Metropolis ``accept-reject" step to the ideal evolution using  $\min\left(1,\exp\left(-\Delta E\right)\right)$ as the accept-reject probability.
For our $80^3$ lattices (which represent the majority of the simulations presented here) the reject probabilities are presented in Table~\ref{tab:idealAR}. The differences imply  that the relative size of the dissipative and real parameters of the pions will weakly depend on $\delta t$.

The downside of having an acceptance probability $p$ different from one is that it introduces a non-trivial renormalization of our time. Effectively, when the ideal step is rejected, the next dissipative step should be thought of as a way to generate a new candidate configuration for the ideal step; the clock freezes. The leading effect of a non-zero rejection probability for the ideal step can then be absorbed by rescaling $\Delta t$ by the acceptance probability $p$. As a result, for all our simulations, the time variable we use is defined as
\begin{align}
  t =  p n_{steps}\Delta t \, ,
\end{align}
with the corresponding $p$ read from Table~\ref{tab:idealAR} and $\Delta t$ is the global timestep of our algorithm.

As the results presented through this work support, this procedure allows us to faithfully correct our time variable. It is nonetheless true that it introduces some uncontrolled subleading systematic errors which may impede us from performing precision measurements in the future. This, together with the fact that the acceptance rate degrades for larger lattices, will lead us to use smaller ideal time steps for future simulations. It may also be worth investigating higher order symplectic integrators, which would help to keep the reject probability small even for large volumes.

\begin{table}
  \centering
\begin{tabular}{|c|c|c|}
  \hline
   $z$ &  $H$ &  Ideal accept probability, $p$\\
  \hline
  3.86978 & 0.003 & 0.958 \\
  \hline
  1.34899 & 0.003 & 0.953 \\
  \hline
  0.94429 & 0.003 & 0.951 \\
  \hline
  0       & 0.002 & 0.948 \\
  \hline
  0       & 0.003 & 0.948 \\
  \hline
  0       & 0.004 & 0.948 \\
  \hline
  0       & 0.006 & 0.947 \\
  \hline
  0       & 0.01  & 0.946 \\
  \hline
  -2.2011 & 0.003 & 0.940 \\
  \hline
\end{tabular}
\caption{\label{tab:idealAR} Accept-reject probability associated to our ideal step for different simulations.}
\end{table}

\subsubsection{Viscous steps for $\phi$}
\label{sec:viscstepphi}

The spatially discretized equation to be solved is
\st
\partial_t \phi_a  =  -\Gamma_0 \frac{\partial {\mathcal H}}{\partial \phi_a}  + \theta_a \, ,
\stp
where the noise correlator is given by a discretized \Eq{eq:langevin_var}.

We will realize the Langevin process  with Metropolis updates.
Briefly,  an update proposal is made for a lattice site $x$
\st
\phi_a(t+ \delta t,x) = \phi_a(t,x) +  \Delta \phi_a \, ,
\stp
where for each flavor index $a$  the increment is
\[
\Delta \phi_a =\sqrt{ 2 \delta t \Gamma_0 } \, \xi_0 \, ,
\]
Here  $\xi_0$ is a random number with unit variance $\llangle \xi^2_0 \rrangle =1$.
In practice $\xi_0$ is generated from a flat distribution between $[-\sqrt{3},\sqrt{3}]$,  since this is faster than generating Gaussian random numbers. The update proposal is
accepted with probability ${\rm min} (1, e^{-\Delta \mathcal H})$, where $\Delta\mathcal H$ is the change in the discretized Hamiltonian. If the proposal is rejected $\phi(t+\delta t,x) = \phi(t, x)$.
For $\Delta \phi$  small,
\st
\Delta \mathcal H \simeq  \left. \frac{ \partial \mathcal H}{\partial \phi} \right|_{\phi_a(x,t)} \Delta \phi_a \, ,
\stp
and then the mean and variance of the accepted proposals reproduce the  dissipative and stochastic terms of the Langevin process:
\begin{subequations}
\begin{align}
\label{eq:average_equation1}
   \overline{\phi_a(t+ \delta t,x)-\phi_a(t,x)}  =& - \delta t\,  \Gamma_0  \frac{\partial \mathcal H}{\partial \phi}+ \mathcal{O}(\delta t^2)  \, ,  \\
      \overline{\left(\phi_a(t+ \delta t,x)-\phi_a(t,x)\right)^2}  =& 2 \delta t\Gamma_0  + \mathcal{O}(\delta t^2)  \ .\label{eq:average_equation2}
\end{align}
\end{subequations}

For the sake of clarity, let us rederive this result.
Consider the Markov process generated by the Metropolis algorithm: $\Delta \phi$ is accepted if $e^{-\Delta \mathcal{H}}-1$ is positive, otherwise it is accepted only with probability $e^{-\Delta \mathcal{H}}$. Employing the step function $\theta(x)$, the update rules for each lattice site can be written as
\begin{subequations}
\begin{align}
\phi_a(t+ \delta t,x) &= \phi_a(t,x) + \theta(e^{-\Delta \mathcal{H}}-1) \, \Delta \phi_a +
\theta(1-e^{-\Delta \mathcal{H}}) \, e^{-\Delta \mathcal{H}} \Delta \phi_a \, , \\
& = \phi_a(t,x) +  \Delta \phi_a +
\theta(1-e^{-\Delta \mathcal{H}}) \, \left(e^{-\Delta \mathcal{H}} -1\right) \Delta \phi_a \ .
\end{align}
\end{subequations}
In the limit of small $\delta t$, $\Delta\phi_a$ is small, and we can Taylor expand the energy and the probability, obtaining
\st
\phi_a(t+ \delta t,x) = \phi_a(t,x) +  \Delta \phi_a  \\
+\theta\left(  \frac{ \partial \mathcal H}{\partial \phi}  \sqrt{2 \delta t \Gamma_0 } \, \xi_0\right) \, \left(
-   \frac{\partial \mathcal H}{\partial \phi} \right)\, 2 \delta t \Gamma_0  \, \xi_0^2  \, .
\stp
Taking averages and noting that the $\theta$-function vanishes for half of the realizations, one immediately reproduces Eqs.~\eqref{eq:average_equation1}-\eqref{eq:average_equation2}.

To iterate over the sites, we loop over the lattice in a checkerboard pattern, first updating all of the even sites, and then updating the odd sites.
Since the interactions are nearest neighbors only, the even site updates are independent of each
other and can be done in any order.  In addition, the checkerboard Metropolis updates maintain the lattice translational invariance
and are easy to implement with PETSc and MPI \cite{petsc-web-page,petsc-efficient}.

 Finally we turn to the step size $\delta t$.
We would
like the computer time required to thermalize modes of wavelength $\sim a$ to be as short as possible.
If $\delta t$ is small, then the steps
are always accepted, but lead only to  a small change in $\phi$;  equilibration then requires many steps.
If $\delta t$ is  large, then $\Delta \phi$ is large,  but the updates are always rejected, again requiring many steps.
We have found that choosing $\delta t = 0.04/\Gamma_0 $ leads to an accept-reject probability of approximately $0.5$, optimizing these considerations.

\subsubsection{Viscous steps for charges $n_A$ and $n_V$}
\label{sec:viscstepcharge}

We are considering the evolution equation $\partial_t U = \mathcal O_C(U)$.
Since each charge in the tensor $n_{ab}$ is independent we will dispense with the flavor indices in
the rest of this section. All the updates described here will be applied
in sequence
to the three axial charges $n_A^s$ and the three vector charges $n_V^s$.
The continuum equation to be solved is the stochastic diffusion equation
\begin{align}
   \label{eq:stochasticdiff}
   \partial_t n + \partial_i j^i = 0 \, ,   \qquad  j^i = -\frac{\sigma_0}{\chi} \partial^i n  + \Xi^i \, ,
\end{align}
and equilibrium effective Hamiltonian is\footnote{Again this action describes only one isospin component of
   the iso-axial or iso-vector charge. In general
   \st
   \mathcal H = \int d^3x  \left(\frac{\vec{n}_A \cdot \vec{n}_A}{2\chi_0} +  \frac{\vec{n}_V \cdot \vec{n}_V}{2\chi_0}  \right)
   = \frac{1}{4\chi_0} \int d^3x \, n_{ab} n_{ab} \, ,
   \stp
   and $n_{ab} n_{ab}$ is written $n^2$ in the majority of the text.
}
\st
 \mathcal H = \int d^3x \, \frac{n^2 }{2\chi_0} \, .
\stp

To generate the Langevin dynamics in \eqref{eq:stochasticdiff} we will again use Metropolis steps.
In order to get the correct diffusive dynamics at long wavelengths
the charge must be exactly conserved by the update proposals. We therefore update
the cells in pairs by making a  Metropolis proposal for the charge transferred
between two cells over a time $\delta t$.

The figure below shows a few sites of the lattice, with the even sites painted grey.
Integrating \eqref{eq:stochasticdiff} over the spatial volume of lattice cell $A$ and time $\delta t$, the discretized
equation of motion for the charge takes the form
\st
\hat n(\hat t + \delta \hat t, \hat x) = \hat n(\hat t, \hat x)  - (Q^x_+ - Q^x_-) - (Q^y_+ - Q^y_-)
-(Q_+^z - Q_-^z) \, ,
\stp
where, for example,  $Q^x_+$  is the charge transfer between  $A$ and $B$
over a time $\delta t$. (For clarity below we have restored the hats to indicate quantities in lattice units, e.g. $\hat n = na^3$)
\begin{center}
   \includegraphics[width=0.4\textwidth]{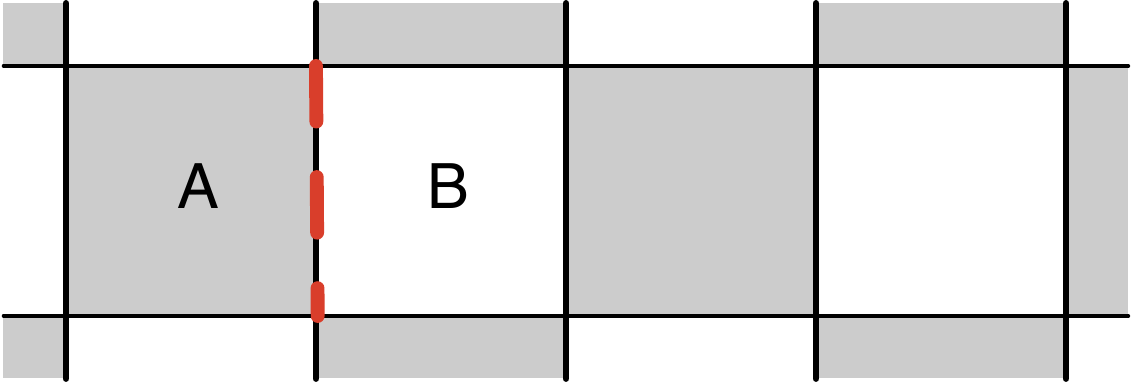}
\end{center}
The proposed  Metropolis flux through the interface is
\st
Q_+^x =  q =  \sqrt{ 2 \hat \sigma_0 \delta \hat t} \, \xi_0 \, ,
\stp
where again $\xi_0$ is a uniform random number with unit variance.
Thus the proposed update for cells $A$ and $B$ is
\begin{align}
   \hat n_A \rightarrow& \hat n_A -q \, ,  \\
   \hat n_B \rightarrow& \hat n_B +q \, ,
 \end{align}
and change in action by the proposed change is
\begin{align}
   \Delta \hat{\mathcal H} =& \frac{(\hat n_B+q)^2}{2\hat \chi_0} +  \frac{(\hat n_A - q)^2 }{2\hat \chi_0} - \frac{\hat n_B^2}{2\hat \chi_0}  - \frac{\hat n_A^2}{2\hat \chi_0} \, ,  \\
   =&  (\hat n_B - \hat n_A) \frac{q}{\hat \chi_0}  + \mathcal O(q^2) \, .
\end{align}
The proposed updated is accepted with probability ${\rm min}(1 , \exp(-\Delta \hat{\mathcal H}) )$.
Then it is easy to see that mean charge transfer is
\begin{align}
   \overline{q} =&  - (\hat n_B - \hat n_A) \frac{\hat \sigma_0}{\hat \chi_0} \delta \hat{t}  \, , \\
\simeq& - a^2 \delta t \, D_0 \partial_x n \, ,
\end{align}
which is the expected charge transfer for a diffusive step.
Finally, is easy to show that the flux
 $\Xi^x \simeq  q/(\delta t a^2)$  has the expected variance.
Thus for small $\delta t$ the Markov updates  produce an equivalent update
to the Langevin step.

To iterate over the faces of the lattice we again divide the cells into a checkerboard pattern.
We first do the Metropolis updates for all of the $x_+$ interfaces for all of the even cells, i.e. cell $A$ is even and cell $B$ is
odd as shown in the figure above. These updates
are independent of each other and can be done in any order. This step is followed by Metropolis updates
of the $x_-$ interfaces of the even cells, i.e. now cell $A$ is odd and cell $B$ is even. Then
we proceed to update the $y$ and $z$ directions in a similar manner.  To eliminate potential bias,
the order of the $(x, y, z)$ iterations and the $(+,-)$ iterations are each randomly shuffled for each iteration of the $C$ stage
of the Markov chain.




\section{Thermodynamics and statics of the model}
\label{sec:thermoapp}





\subsection{Fixing the critical temperature}
\label{sec:fixingTc}
\begin{figure}
 \centering
 \includegraphics[width=0.49\textwidth]{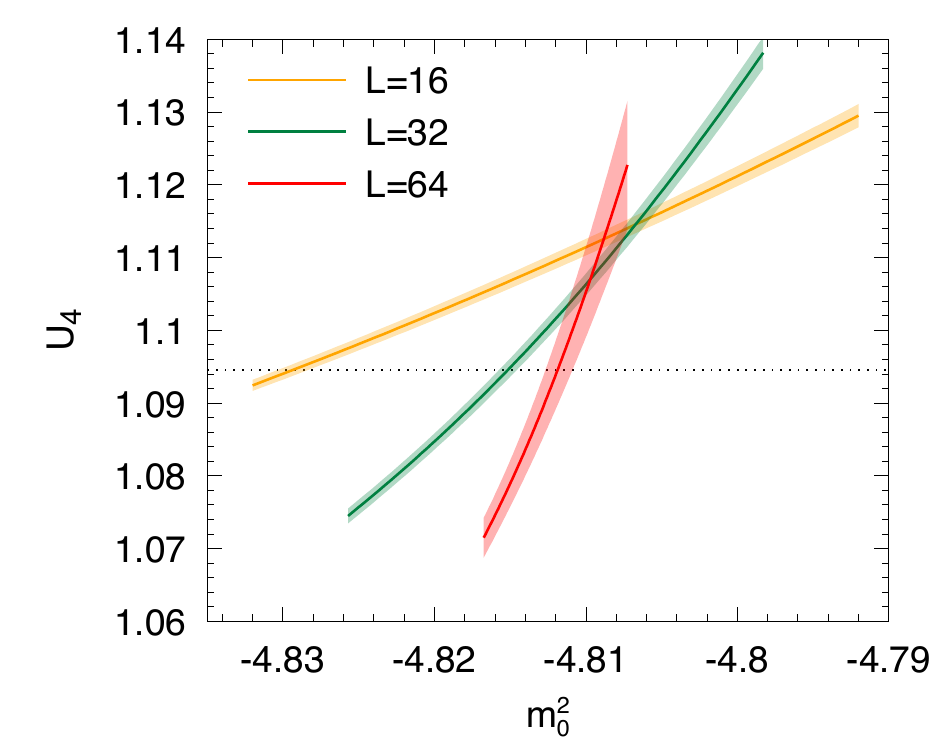}
 \includegraphics[width=0.49\textwidth]{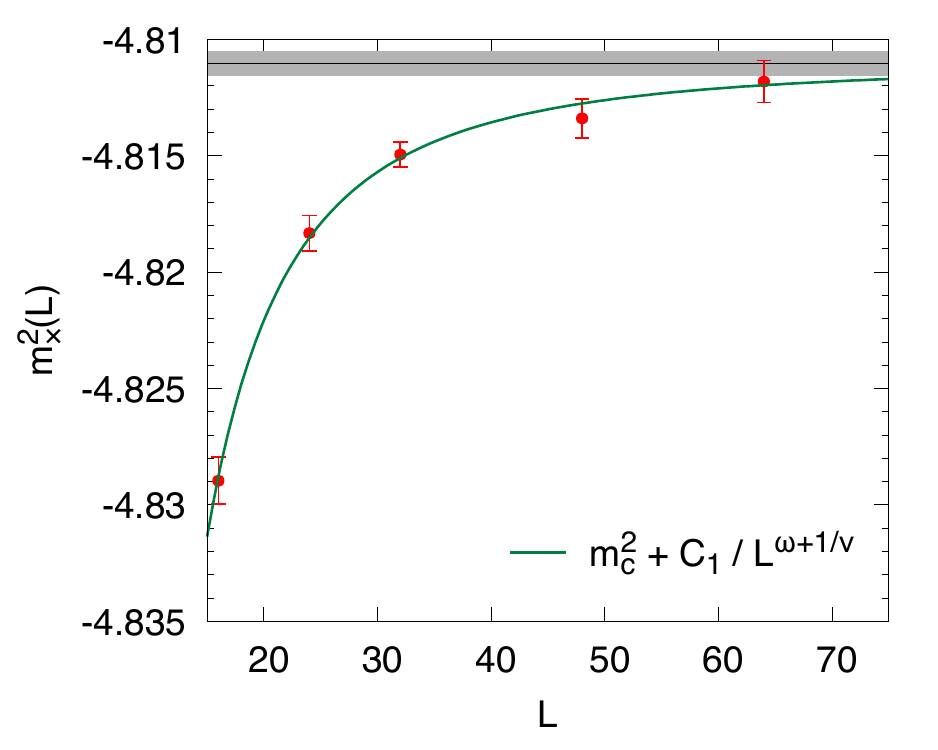}
 \caption{\label{fig:bindercm2}
    \noindent {\bf Left:} The Binder cumulant $U_4$ as a function of $m^2_0$ and  $L$. For clarity, the $L=24$ and $L=48$ results are not shown.
    The asymptotic value of the Binder cumulant $U_4^c$ is shown as the dotted line, and is taken from \cite{Hasenbusch:2000ph}.
 The crossing points, $m^2_\times(L)$, are when bands cross the dotted line.  {\bf Right:}  A fit to the Binder crossing formula in \eqref{eq:m2times} which determine the critical parameter $m^2_c=-4.8110(4)$.  }
\end{figure}

In this appendix we will describe our (conventional) strategy for locating the critical point of the model, by adjusting the bare coupling $m^2_0$. A good summary of the technique is given in \cite{Arnold:2001ir}. Throughout this section we set $\lambda=4$ (somewhat arbitrarily) and $H=0$. After a preliminary search in $m^2_0$, we ran a set of long simulations at $m^2_0 = -4.812$  for $N=16,24,32,48,64$.
For each of these simulations, we  used  reweighted samples to compute the Binder
cumulant \cite{Binder:1986uh} for a range of $m^2_0$. More explicitly we computed
\begin{align}
   \llangle M^2    \rrangle =& \frac{ \sum_t e^{-\frac{1}{2} \delta m^2_0  \sum_{x} \phi^2(t,x) }  M_a(t) \cdot M_a(t)  }{ \sum_{t} e^{-\frac{1}{2} \delta m^2_0 \sum_x \phi^2(t,x) } }  \, , \\
   \llangle (M^2)^2    \rrangle  =&  \frac{ \sum_t e^{-\frac{1}{2} \delta m^2_0  \sum_x \phi^2(t,x) }  (M_a(t) \cdot M_a(t))^2  }{ \sum_{t} e^{-\frac{1}{2} \delta m^2_0 \sum_x\phi^2(t,x) } } \, ,
\end{align}
and then determine the Binder cumulant
\st
U_{4} \equiv \frac{  \llangle (M^2)^2    \rrangle }{\llangle M^2    \rrangle^2 } \, .
\stp
A plot of
$U_4$ for our  $N=16,32,64$ samples is shown in the left panel of \Fig{fig:bindercm2}.

In the high temperature phase, the fluctuations of the order parameter are Gaussian
and the Binder cumulant reaches $(N+2)/N=1.5$, while in
the low temperature phase the system is ordered and the Binder
cumulant is unity.
In the critical region (where the system size is of order the correlation length)
the Binder cumulant transitions between $1$ and $1.5$, and approaches a universal value $U_4^c$ at the
critical temperature
for large $L$~\cite{Springer:2015kxa}.
From a precision study by Hasenbusch~\cite{Hasenbusch:2000ph}, we have taken this asymptotic value to be  $U_{4}^c=1.0945(1)$.  Now, for each $L$,  we  determined a value of $m^2_0$,  denoted  $m^2_\times(L)$,  where the Binder cumulant reaches $U_4^c$ (see figure). For large
$L$ the expected scaling of  $m^2_\times(L) - m^2_c$ is
\st
\label{eq:m2times}
m^2_\times(L) - m^2_c = \frac{C_1 }{L^{1/\nu + \omega}  } \, ,
\stp
and this scaling can be used to determine $m_c^2$.
Then we fit $m^2_\times(L)$, with the functional form  in \eqref{eq:m2times}
to find the value of $m_c^2$ quoted in the body of the text in \eqref{eq:mc2}.
A plot of our fit is shown in the right panel of \Fig{fig:bindercm2} and has $\chi^2/{\rm dof}=0.5$, suggesting that the error bars have been slightly overestimated.

\subsection{Thermodynamics on the critical line}
\label{sec:thermoh}

In this section we continue the discussion in \Sect{sec:thermo},
and determine the non-universal parameter $H_0$ by making a scan on the critical line.

In practice, we made a scan only  approximately on the critical line at $m^2=-4.8130$,  and then used reweighting to determine $\bar{\sigma}$ at our nominal value of $m^2_c =-4.8110$.
In the scaling theory the condensate takes the form
\st
  \bar{\sigma} =  h^{1/\delta} (f_G(z, z_L)  + h^{\omega \nu_c}  f_G^{(1)} (z,z_L)) \, .
\stp
where in addition to the scaling variables $h$ and $z$,
we have included an additional dependence on the system size $L$ through the  scaling variable $z_L = L_0/L h^{\nu_c}$,
and its associated constant $L_0$~\cite{Engels:2014bra}. The scaling function $f_G(z, z_L)$ has been parameterized for $z_L = [0, 1.2]$ by Engels and Karsch~\cite{Engels:2014bra} (see their Eq. 29). We have  also included a subleading scaling function, $f_{G}^{(1)}(z,z_L)$, which provides a correction
to the leading asymptotics close to $T_c$.  We are working on the critical line where $z=0$, and the dependence on $z_L$ is weak for $L$ large. Thus we will neglect the dependence on $z_L$ in the subleading term and describe our data with the form
%
\st
\label{eq:finitevolumefit}
\bar{\sigma} =  h^{1/\delta} (f_G(0, z_L) +   h^{\omega \nu_c} C_H ) \, .
\stp
For the critical exponents here and below
we  use the results from~\cite{Engels:2014bra}
\st
\beta = 0.380(2)\,, \qquad \delta = 0.4824(9) \, ,
\stp
and then used the hyperscaling relations to determine all others, e.g.
  $d \nu = \beta (1 + \delta) \simeq 2.213$.
We have taken $\omega = 0.77$ for the subleading exponent from \cite{Hasenbusch:2000ph}.
The data for $\bar{\sigma}$ on $32^3$ and $64^3$ lattices on the critical line are shown in the left panel of \Fig{fig:hcrit} in the body of the text. They were fit in the range
$z_L=[0,1.2]$ with \eqref{eq:finitevolumefit},
which fixes the three fit parameters for $H_0$, $L_0$, and $C_H$:
   \st
   \label{eq:finitevolfitresults}
   H_0 = 5.15(15), \qquad L_0 = 0.97(4), \quad \mbox{and} \quad C_H = 0.54(4) \, ,
   \stp
   with $\chi^2/{\rm dof} = 2$. $H_0$ is recorded in Eq.~\eqref{eq:mc2}.
Also shown in the left panel of \Fig{fig:hcrit} is the predicted magnetization from the fit
at infinite volume (the dashed line). We see that already at $L=64$, we are  essentially at  infinite volume for the range of $H$ considered in this work.
Indeed a simple two parameter fit to our $L=64$ results (not shown) with a simple form,
$\bar{\sigma} = (H/H_0)^{1/\delta} \left(1 + C_H (H/H_0)^{\omega \nu_c} \right)$,
yields compatible results for $H_0$.

\subsection{Thermodynamics at $H=0$}
\label{sec:ThermoT}

In this section we continue the discussion in \Sect{sec:thermo},
and determine the non-universal parameter $\mm^2$ (or the amplitude $B$) by making a scan at $H=0$.  By measuring $\llangle M^2 \rrangle$ we will extract the condensate, $\Sigma$, at infinite volume and zero field  defined in \eqref{eq:SigmaDef}.

The leading deviation of $\llangle M^2 \rrangle$ from $\Sigma^2$ at finite volume comes from the fluctuations of long wavelength Goldstone modes, and can be neatly analyzed with a Euclidean pion effective theory~\cite{Hasenfratz:1989pk}, which was briefly discussed in \Sect{sec:thermoapppioneft}.
 The resulting
expansion relating $\llangle M^2 \rrangle $ and $\Sigma^2$ is only for  $f^2(T) L \gg 1$, and takes the form
\st
\label{eq:hasenfratz}
\llangle M^2 \rrangle = \Sigma^2 \left(\rho^2_1 + \frac{8\rho_2}{f^4 L^2} \right) + \mathcal O((f^2 L)^{-3}) \, .
\stp
Here the symmetry group $O(N)$ is broken to $O(N-1)$,  $\rho_1$ and $\rho_2$ are expansions in $1/L$
\begin{align}
   \rho_1 &= 1 + \frac{ (N-1)\beta_1}{2 f^2 L } - \frac{(N-1)(N-3)}{8 f^4 L^2 } (\beta_1^2 - 2 \beta_2 ) \, , \\
\rho_2 &= \frac{(N-1) \beta_2}{4}  \, ,
\end{align}
with the shape coefficients that record specific sums over the discretized Fourier modes of a cubic box of length $L$:
\st
   \beta_1 = 0.225785\, , \qquad \beta_2 = 0.010608 \, .
   \stp
Substituting the shape coefficients, setting $N=4$,  and finally expanding in $1/L$ yields the expansion we will use
\st
 \label{eq:hasenfratznumeric}
 \llangle M^2 \rrangle = \Sigma^2 \left[  1   +\frac{0.677355}{f^2 L} + \frac{0.156028}{f^4 L^2}  + O\left((f^2L)^{-3}\right) \right] \, .
\stp

To extract $\Sigma(T)$ we performed a sequence of simulations at $L=16,24,32,48,64$ at $H=0$. We then used \eqref{eq:hasenfratz}
with $N=4$ to  fit $f^2(T)$ and $\Sigma(T)$ using the $L=32$, $48$, and $64$ points (not shown). The difference in the first and second orders in the expansion was
used to estimate the systematic uncertainty in the extracted values of $f^2$ and $\Sigma$. For $\Sigma(T)$ this is smaller
than our statistical uncertainty, which would not have been the case if  only the leading term $1/L$ term in the expansion \eqref{eq:hasenfratznumeric} had
been used.  We also found that with the quadratic term provides a better description of  the data with no additional parameters,
providing credance to the pion EFT in this range of temperatures and volumes.

\subsection{Extracting the pion's decay constant and screening mass}
\label{sec:thermoapppioneft}

\begin{figure}
   \centering
   \includegraphics[width=\textwidth]{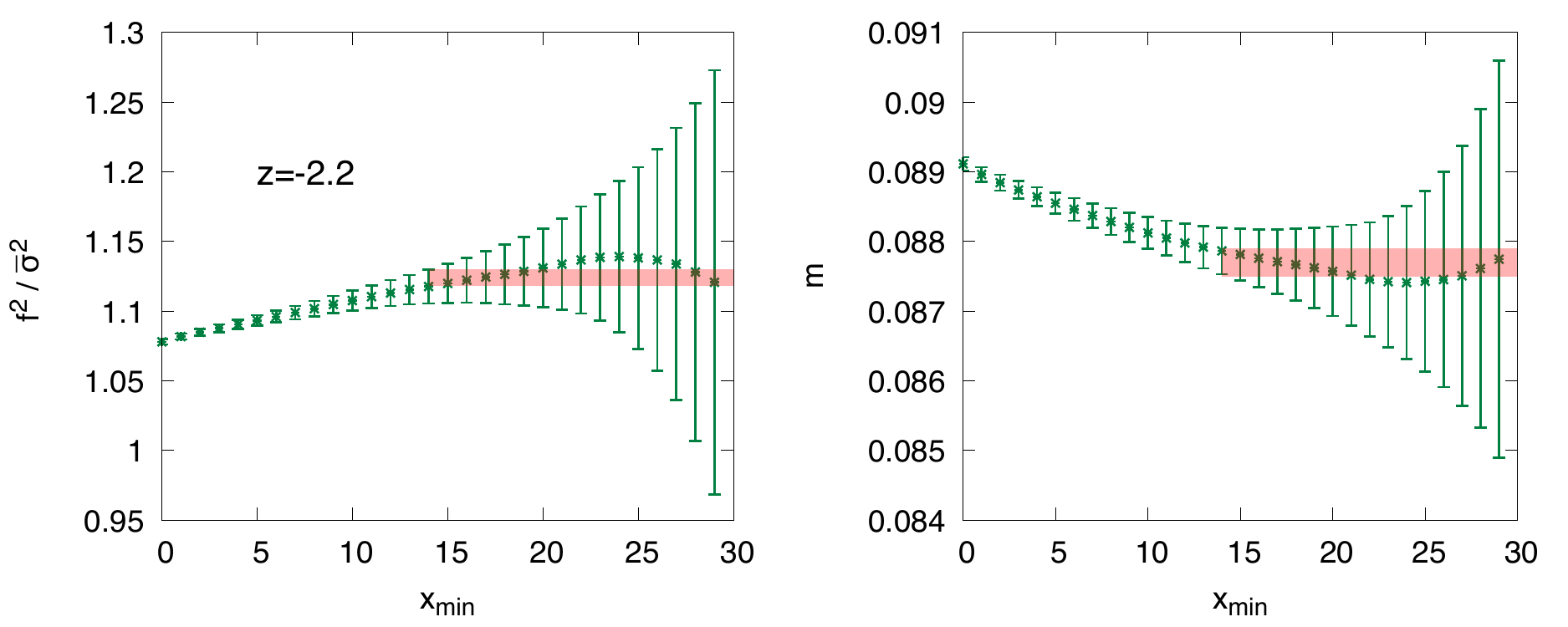}
   \caption{\label{fig:staticfandm} Pion decay constant and static screening mass extracted from a single state fit of the pion wall-to-wall static correlator $D(x)$. To remove the contamination from higher states, we restrict our fit (not shown) to the range $[x_{\rm min}, N-x_{\rm min}-1]$. We plot here the fitted parameters as a function of $x_{\rm min}$. Our nominal values are shown in green and computed by averaging the results for $x_{\rm min}>15$; once the values have plateau-ed.}
\end{figure}

The aim of this section is to verify the static part of the  Gell-Mann-Oakes-Renner (GOR) formula
\begin{align}
   f^2 m^2 = H\bar\sigma \ ,
\end{align}
which is discussed in \Sect{sec:stateft}.
To this end, we return to our simulation in the broken phase and perform some more static measurements.
The magnetization is straightforward to measure
\begin{align}
   \bar\sigma=0.34906\pm0.00003 \, ({\rm stat.}) \, ,
\end{align}
and gives for right-hand side
\begin{align}
   H\bar\sigma  = (1.04718 \pm 0.00009\, ({\rm stat.}))\cdot10^{-3}\label{eq:res_mpstatApp} \ .
\end{align}
To measure the pion decay constant $f^2$ and the screening mass $m$, we follow \cite{Engels:2009tv} and fit (not shown) the static pion wall-to-wall static
correlator
\begin{align}
  D(x) \equiv \frac{1}{3L^2}\sum_s\sum_{y,y',z,z'}\langle \pi_s (t=0, x, y, z)\pi_s (t=0, x=0, y', z')\rangle
\end{align}
to a single state ansatz in a periodic box of size $L$
\begin{align}
  D(x)= \frac{\bar\sigma^2}{2m f^2}\frac{e^{-mx}+e^{-m(L-x)}}{1 - e^{-mL}}\, .
\end{align}
To reduce the remaining effects of higher states, we reduce the range of our fit to $[x_{\rm min},L-x_{\rm min}-1]$ and study the dependence of the parameters on $x_{\rm min}$. Their nominal value is then extracted by fitting the resulting plateaus. This procedure is illustrated in Fig.~\ref{fig:staticfandm} and lead to the following determination
\begin{align}
   \frac{f^2}{\bar\sigma^2} &= 1.124\pm0.006\, ({\rm stat.}) \, ,\\
   m &= (8.77\pm0.02\, ({\rm stat.}))\cdot 10^{-2} \ ,
\end{align}
leading to
\begin{align}
   f^2 m^2 =(1.053 \pm 0.007\,({\rm stat.})) \cdot 10^{-3} \, .  
\end{align}
We see that the static GOR relation is satisfied within statistical errors, as displayed in \eqref{eq:GOR1} in the body of the text.





%

\end{document}